\documentclass[11pt]{article}
\usepackage{amssymb,amsmath,amsfonts}
\usepackage{graphicx}
\usepackage{graphics}
\usepackage{eepic,epsfig}
\usepackage{verbatim}
\usepackage{color}
\usepackage{hyperref}
1.60cm \makeatletter \@addtoreset{equation}{section}

\makeatother

\begin{document}
\title{Induced current in high-dimensional AdS spacetime in the presence of a cosmic string and a compactified extra dimension}
\author{W. Oliveira dos Santos\thanks{E-mail: wagner.physics@gmail.com}, H. F. Mota\thanks{E-mail: hmota@fisica.ufpb.br} and  E. R. Bezerra de Mello\thanks
	{E-mail: emello@fisica.ufpb.br}.\\
\\
\textit{Departamento de F\'{\i}sica, Universidade Federal da Para\'{\i}ba}\\
\textit{58059-900, Caixa Postal 5008, Jo\~{a}o Pessoa, PB, Brazil.}\vspace{%
0.3cm}\\
}
\maketitle
%
\begin{abstract}
In this paper, we analyse the bosonic current densities induced by a 
magnetic flux running along the core of an idealized cosmic string in a
high-dimensional AdS spacetime, admitting that an extra dimension coordinate is compactified. Additionally we admit the presence of a
magnetic flux enclosed by the compactified axis. 
In order to develop this analysis we calculate
the complete set of normalized bosonic wave-functions obeying a
quasiperiodicity condition, with arbitrary phase $\beta$, along the compactified extra dimension. In this context, 
only azimuthal and axial currents densities take place. As to the azimuthal current, two contributions appear. 
 The first one corresponds to the standard azimuthal
current in high-dimensional AdS spacetime with a cosmic string without compactification while the
second contribution is a new one, induced by the
compactification itself. The latter is an even
function of the magnetic flux enclosed by the compactified axis and is an odd function 
of the magnetic flux along its core with period equal to quantum flux, $\Phi_0=2\pi/e$.
On the other hand, the 
nonzero axial current density is an even function 
of the magnetic flux along the core of the string and an odd function 
of the magnetic flux enclosed by the compactified axis. We also find that the
axial current density vanishes for untwisted and twisted bosonic
fields in the absence of the magnetic flux enclosed by the compactified axis.
Some asymptotic expressions for the current density are provided for
specific  limiting cases of the physical 
parameter of the model.

\end{abstract}
\bigskip
PACS numbers: 03.70.+k 04.62.+v 04.20.Gz 11.27.+d\\
\bigskip
%
\section{Introduction}
\label{Int}
%
\hspace{0.55cm}The physics underlying quantum vacuum fluctuations arises once quantum aspects of relativistic phenomena are taken into account. That means a quantized relativistic field (scalar, electromagnetic or fermionic) will have a fluctuating ground state. In Minkowski spacetime, for instance, the Vacuum Expectation Value (VEV) of physical observables, as a consequence of quantum vacuum fluctuations of relativistic fields, is zero unless the vacuum is somehow `perturbed' by external influences. These external influences are in general boundary conditions of some sort or coupled external fields. One very known physical observable that gets a nonzero VEV under external influences is the energy density that characterizes the Casimir effect \cite{Mostepanenko:1997sw, bordag2009advances, Milton:2001yy}. Another physical observable of interest that averages to a nonzero value under these circunstances is the four-current density due to charged fields. This is of special importance since the VEV of the four-current density can provide a better understanding of the dynamics of the electromagnetic field once it is used as a source in the semiclassical Maxwell equations.

An additional feature related to modifications of quantum vacuum fluctuations of relativistic fields is its occurrence face a curved background. It has been known that geometrical and topological aspects of a curved spacetime also induce a nonzero VEV of physical observables \cite{Mostepanenko:1997sw, bordag2009advances, Milton:2001yy}. In particular, the induced VEV of the four-current density by curved backgrounds has been investigated in Refs. \cite{Braganca:2014qma, deMello:2014hya, deMello:2014ksa, deMello:2013loa, BezerradeMello:2009km, BezerradeMello:2010ii, Mohammadi:2014oga, deSousa:2015bvy}. Among these curved backgrounds, the anti-de Sitter (AdS) spacetime carries very interesting properties which provide strong motivations to study it \cite{Sokolowski:2016tar, deMello:2011ji, Santos:2015mja, deMello:2015yda,Bellucci1,Bellucci2}.

By considering a negative cosmological constant, the AdS spacetime is obtained as a solution of the Einstein's equations and thus, is characterized by a constant negative curvature. Thereby, from a theoretical and fundamental point of view, the AdS spacetime makes possible several problems to be solved exactly as a consequence of its high symmetry, allowing the quantization of fields more easily, besides offering better insights into the quantization of fields in other curved spacetimes. Moreover, the AdS spacetime arises as a ground state solution of string and supergravity theories and also appears in the context of AdS/CFT correspondence, scenario which makes possible the realization of the holographic principle, relating string theory (supergravity) in a high-dimensional AdS spacetime with a conformal field theory constructed in its boundary \cite{Aharony:1999ti}. In addition, the AdS background geometry is relevant in branewold scenarios with large extra dimensions, offering a way to solve the hierarchy problem between the gravitational and electroweak mass scales \cite{Brax:2003fv}.

The combination of the AdS spacetime with the spacetime of a cosmic string is even more interesting, since this combined geometry  makes possible to identify in the VEV of some observable, the contributions come from either parts, namely, from the AdS geometry and cosmic string topology. Cosmic strings are linear topological defects that are predicted in the context of both gauge field theories and supersymmetric extensions of the Standard Model of particle physics, as well as in the context of string theory \cite{VS, hindmarsh, escidoc:153364, Copeland:2011dx, Hindmarsh:2011qj, Chernoff:2017pui}. The spacetime of a straight, infinitely long and structureless cosmic string  is characterized by a conical topology arising due to the angle deficit in the plane perpendicular to it \cite{VS, hindmarsh, escidoc:153364}. Phenomenologically, current observations of CMB suggest cosmic strings can contribute to a small percentage of the primordial density perturbations \cite{Ade:2013xla} in the universe and can also play a important role in other cosmological, astrophysical and gravitational phenomena \cite{escidoc:153364, Copeland:2011dx, Hindmarsh:2011qj, Chernoff:2017pui}.

In the present paper we are interested in calculating the VEV of current density associated with a charged scalar field, and investigate the effects arising from the geometry and topology of a high-dimensional AdS spacetime in the presence of a cosmic string carrying a magnetic flux. In addition we will assume a compactification of one extra dimension and the existence of a constant vector potential along it.  Thus, the presence of these magnetic fluxes as well as the compactified extra dimension will also provide additional contributions to the VEV of the current density, as we shall see.

The presence of extra compact dimensions is a characteristic aspect of all the theories mentioned before where the AdS spacetime plays a key role and, as previously said, induce nonzero contributions to physical observables such as the energy-momentum tensor which has not only the energy density component but also the stresses components (see \cite{deMello:2014hya} and references therein). In this case, for instance, the vacuum energy density induced by the extra compact dimensions offers an explanation for the observed and still unexplained accelerated expansion of the universe. In Kaluza-Klein-type models and in braneworld scenarious, on the other hand, the dependence of the size of the compact extra dimension by the vacuum energy density serves as mechanism to stabilize fields known as moduli fields.

This paper is organized as follows. In section \ref{sec2} we present the high-dimensional AdS spacetime in the presence of a cosmic string and obtain the complete set of normalized solutions of the Klein-Gordon equation associated with a charged scalar field in this background, considering the presence of a azimuthal and axial vector potentials . This solution is then submitted to a nontrivial boundary condition that compactifies an extra dimension. This set of solution  is used to construct the Wightman function. In section \ref{sec3} we first prove that the VEV's of the charge density, radial current density and current density associated with the extra dimensions, except the one that is compactifed, are all zero. Finally, the rest of the section 3 is devoted to compute the nonzero azimuthal current density and the nonzero current density associated with the compactified extra dimension. In this case, we show that the azimuthal current density has a pure contribution due to the high-dimensional AdS spacetime with a cosmic string plus a second contribution due to the compactification of the extra dimension. Moreover, we also show that the current density associated with the compactified extra dimension has only the contribution due to the compactification. The section \ref{sec4} is devoted to the main 
conclusions about our results. Throughout the paper we use natural units $G=\hbar = c = 1$.

\section{Klein-Gordon equation and Wightman function}
\label{sec2}

The main objective of this section is to obtain the positive frequency Wightman function associated with a massive scalar field in a $(D+1)$-dimensional AdS spacetime, with $D>3$, in presence of a cosmic string and a compactified extra dimension. This function is important in the calculation of vacuum polarization effects. In
order to do that we first obtain the complete set of normalized mode
functions for the Klein-Gordon equation admitting an arbitrary curvature
coupling parameter.

In cylindrical coordinates, the geometry associated with a cosmic string in
a $(3+1)$-dimensional AdS spacetime is given by the line element below
(considering a static string along the $y$-axis):
\begin{equation}
ds^{2}=e^{-2y/a}[dt^{2}-dr^{2}-r^{2}d\phi ^{2}]-dy^{2}\ ,  \label{ds1}
\end{equation}
where $r\geqslant 0$ and $\phi \in \lbrack 0,\ 2\pi /q]$ define the
coordinates on the conical geometry, $(t, \ y)\in (-\infty ,\ \infty )$, and
the parameter $a$ determines the curvature scale of the background
spacetime. The parameter $q\geq 1$ codifies the presence of the
cosmic string. Using the \textit{Poincar\'{e}} coordinate defined by $w=ae^{y/a}$, the line element above is written in the form conformally related to the line element associated with a cosmic string in Minkowski spacetime:
\begin{equation}
ds^2 = \left(\frac{a}{w}\right)^2[dt^2 - dr^2 - r^2d\phi^2 - dw^2 ] 
\label{ds2}
\end{equation}
For the new coordinate one has $w\in \lbrack 0,\ \infty )$. Specific values for this coordinates deserve to be mentioned: $w=0$ and $w=\infty $ correspond to the AdS boundary and horizon, respectively.

For an idealized cosmic string, i.e., an infinitely thin and long straight cosmic string in the background of Minkowski spacetime, the line element expression inside the brackets of the right-hand
side of (\ref{ds2}), has been derived in \cite{Vile81} by making use of two
approximations: the weak-field approximation and the thin-string one. In
this case the parameter $q$ is related to the mass per unit length $\mu $ of
the string by the formula $1/q=1-4G\mu $, where $G$ is the Newton's
gravitational constant.  However, the validity of the line element with the planar angle deficit has been extended beyond linear perturbation theory by \cite{Gott85,VS}. In this case the parameter $q$ need not to be close to 1. Note that in braneworld scenarios based on AdS spacetime, to which the results given in this paper could be applied, the fundamental Planck scale is much smaller than $m_{\mathrm{Pl}}$ and can be of order
of string formation energy scale.

The generalization of (\ref{ds2}) to $(D+1)$-dimensional AdS spacetimes is done
in the usual way, by adding extra Euclidean coordinates \cite{deMello:2011ji}:
\begin{equation}
\label{HDCS}
ds^2 = \left(\frac{a}{w}\right)^2\bigg[dt^2 - dr^2 - r^2d\phi^2 - dw^2 - \sum_{i=4}^{D}(dx^i)^2\bigg]   \   .
\end{equation}
The Euclidean version of the line element expressed inside the bracket of the above equation, has been presented in \cite{Linet}, and called as {\it conical-type line singularity} in arbitrary dimension; therefore, we consider  the line element inside \eqref{HDCS} as a Minkowski version of the cosmic string metric spacetime for higher-dimension. Moreover, a discussion about the generalization of the cosmic string spacetime can also be found in \cite{Sitenko}.
   
 The curvature scale $a$ in \eqref{HDCS} is related to the cosmological constant, $\Lambda $, and the Ricci scalar, $R$, by the formulas
\begin{equation}
\Lambda =-\frac{D(D-1)}{2a^{2}} \ ,\ \ R=-\frac{D(D+1)}{a^{2}}\ .
\label{LamR}
\end{equation}

The analysis of induced  current density for a charged massive scalar field in the anti-de Sitter (AdS) space described in Poincar\'e coordinates with toroidally compact dimensions, has been developed in \cite{deMello:2014hya}. In the latter it is assumed that, in addition to compact dimensions, the field obeys periodicity conditions with general phases. Moreover, the presence of constant vector potentials has also been considered.

In this present paper we are interested in calculating the induced vacuum current density, $\langle j_{\mu}\rangle$, associated with a charged scalar quantum field, $\varphi (x)$, in the cosmic string spacetime in the AdS bulk induced by the presence of magnetic flux running along the string's core. Moreover, we also assume the compactification along only one extra coordinate, defined by $z$ in the expression below, 
\begin{equation}
ds^2 = \left(\frac{a}{w}\right)^2\bigg[dt^2 - dr^2 - r^2d\phi^2 - dw^2 -dz^2- \sum_{i=5}^{D}(dx^i)^2\bigg]   \   .
\label{le2}
\end{equation}
Note that we will also consider the presence of a constant vector potential along the extra compact dimension. This compactification is implemented by assuming that $z\in[0, \ L]$, and the matter field obeys the quasiperiodicity condition below,
\begin{equation}
\varphi(t,r,\phi, w, z + L, x^5,...,x^{D}) = e^{2\pi i\beta}\varphi(t,r,\phi, w, z, x^5,...,x^{D}),
\label{QPC}
\end{equation}
where $0\leq\beta\leq 1$.  The special cases $\beta =0$ and $\beta =1/2$ correspond to the untwisted and twisted fields, respectively, along the $z$-direction. 

The field equation which governs the quantum dynamics of a charged
bosonic field with mass $m$, in a curved background and in the presence of an 
electromagnetic potential vector, $A_\mu$, reads
\begin{equation}
(g^{\mu\nu}D_{\mu}D_{\nu} + m^2 + \xi R)\varphi(x) = 0  \   , 
\label{KGE}
\end{equation}
being $D_{\mu}=\nabla_{\mu}+ieA_{\mu}$. In addition, we have considered the presence of a non-minimal coupling, $\xi$, between the field and the geometry represented by the Ricci scalar, $R$. Two specific values for the curvature coupling are $\xi = 0$ and $\xi = \frac{D - 1}{4D}$, that correspond to minimal and conformal coupling, respectively. Also we shall assume the existence of the following constant vector potentials,
\begin{equation}
A_{\mu} = (0,0,A_{\phi}, 0, A_z, 0,...,0)  \  ,
\label{VP}
\end{equation}
with $A_{\phi}=-q\Phi_\phi/(2\pi)$ and $A_{z}=-\Phi_z/L$, being $\Phi_\phi$ and $\Phi_z$
the corresponding magnetic fluxes. In quantum field  theory the  condition
\eqref{QPC} changes the spectrum of the vacuum fluctuations compared with the case of uncompactified dimension and, as a consequence, the induced vacuum current density changes. 

In the spacetime defined by \eqref{le2} and in the presence of the vector 
potentials given above, the equation \eqref{KGE} becomes
\begin{eqnarray}
\left[\frac{\partial^2}{\partial t^2} - \frac{\partial^2}{\partial r^2} - \frac{1}{r}\frac{\partial}{\partial r} - \frac{1}{r^2}\left(\frac{\partial}{\partial\phi} + ieA_{\phi}\right)^2 -
\left(\frac{\partial}{\partial z} + ieA_{z}\right)^2\right.\nonumber\\
\left.
- \frac{\partial^2}{\partial w^2}-\frac{(1-D)}{w}\frac{\partial}{\partial w} + \frac{M(D,m,\xi)}{w^2} - \sum_{i=5}^{D}\frac{\partial^2}{\partial (x^i)^2} \right]\varphi(x) = 0  \  . 
\label{KGE2}
\end{eqnarray}
where $M(D,m,\xi) = a^2m^2 - \xi D(D+1)$. 

The equation above is completely separable and its positive energy and regular solution at origin is given by,
\begin{equation}
\varphi(x) = Cw^{\frac{D}{2}}J_{\nu}(pw)J_{q|n +\alpha|}(\lambda r)e^{-iE t + iqn\phi + ik_{z}z + i\vec{k}\cdot\vec{x}_{\parallel}}.
\label{Solu1}
\end{equation}
In the expression above $\vec{x}_{\parallel}$ represents the coordinates along the $(D-4)$ extra dimensions, and $\vec{k}$ the corresponding momentum. Moreover,
\begin{eqnarray}
\nu &=& \sqrt{\frac{D^2}{4} + a^2m^2 - \xi D(D+1)},\nonumber\\
E &=& \sqrt{\lambda^2 + p^2 + \vec{k}^2 + (k_{z} + eA_z)^2},\nonumber\\
\alpha &=& \frac{eA_{\phi}}{q} = -\frac{\Phi_{\phi}}{\Phi_0}.
\label{const}
\end{eqnarray}
being $\Phi_0=\frac{2\pi}{e}$, the quantum flux. In \eqref{Solu1} $J_\mu(z)$ represents the Bessel function \cite{Abra}. 

The quasiperiodicity condition \eqref{QPC} provides a discretization  of the quantum number $k_z$ as shown below:
\begin{equation}
k_z = k_l = \frac{2\pi}{L}(l + \beta), \qquad \text{with}\qquad l = 0,\pm 1, \pm2,...\;.
\label{momentum}
\end{equation}
Therefore 
\begin{eqnarray}
E=E_{l} = \sqrt{\lambda^2 + p^2 + \vec{k}^2 + \tilde{k}_l^2},
\label{const2}
\end{eqnarray}
where 
\begin{eqnarray}
\tilde{k}_l &=& \frac{2\pi}{L}(l + \tilde{\beta}),\nonumber\\
\tilde{\beta} &=& \beta + \frac{eA_zL}{2\pi} = \beta - \frac{\Phi_z}{\Phi_0}.
\label{const3}
\end{eqnarray}

The constant $C$ in \eqref{Solu1} can be obtained by the normalization condition below, 
\begin{eqnarray}
\int d^Dx\sqrt{|g|}g^{00}\varphi_{\sigma'}^{*}(x)\varphi_{\sigma}(x)= \frac{1}{2E}\delta_{\sigma,\sigma'}  \   ,
\label{NC}
\end{eqnarray}
where the delta symbol on the right-hand side is understood as Dirac delta 
function for the continuous quantum number, $\lambda$, $p$ and ${\vec{k}}$, and
Kronecker delta for the discrete ones, $n$ and $k_l$. From \eqref{NC} one finds 
\begin{eqnarray}
|C|= \sqrt{\frac{qa^{1-D}\lambda p}{2E (2\pi)^{D-3}L}}.
\label{NC3}
\end{eqnarray}

So, the normalized bosonic wave-function reads,
\begin{equation}
\varphi_{\sigma}(x) =  \sqrt{\frac{qa^{1-D}\lambda p}{2E (2\pi)^{D-3}L}}w^{\frac{D}{2}}J_{\nu}(pw)J_{q|n +\alpha|}(\lambda r)e^{-iE_{l} t + iqn\phi + ik_{l}z + i\vec{k}\cdot\vec{x}_{\parallel}}  \  .
\label{COS}
\end{equation}

The properties of the vacuum state can be given by the positive 
frequency Wightman function, $W(x,x')=\left\langle 0|\hat{\varphi}(x) \hat{\varphi}^{*}(x')|0 \right\rangle$, where $|0 \rangle$ stands for the vacuum state with respect to the observer placed at rest with respect to the string. To evaluate it we use the mode sum formula below,
\begin{equation}
W(x,x') = \sum_{\sigma}\varphi_{\sigma}(x)\varphi_{\sigma}^{*}(x')  \   .
\label{wight}
\end{equation}
Substituting \eqref{COS} into \eqref{wight} we obtain,
\begin{eqnarray}
W(x,x')& =& \frac{qa^{1-D}(ww')^{\frac{D}{2}}}{2(2\pi)^{D-3}L} \sum_{n=-\infty}^{\infty}e^{inq\Delta\phi}\sum_{l=-\infty}^{\infty}\int d\vec{k}\int_0^{\infty}dpp\int_0^{\infty}d\lambda\lambda\nonumber\\
&\times& J_{q|n +\alpha|}(\lambda r)J_{q|n +\alpha|}(\lambda r')J_{\nu}(pw)J_{\nu}(pw')\frac{e^{-iE_{l}\Delta t+ ik_{l}\Delta z + i\vec{k}\cdot\Delta\vec{x}_{\parallel}}}{E_{l}}  \  ,
\label{wight2}
\end{eqnarray}
where $\Delta t=t-t', \Delta \phi=\phi-\phi', \Delta z=z-z'$ and $\Delta \vec{x}_{\parallel}= \vec{x}_{\parallel}-\vec{x}_{\parallel}'$.

In order to develop the summation over the quantum number $l$ we shall apply
the Abel-Plana summation formula \cite{Saharian2010}, which is given by
\begin{eqnarray}
\sum_{l=-\infty}^{\infty}g(l+\tilde{\beta})f(|l+\tilde
{\beta}|)&=&\int_{0}^{\infty}du[g(u)+g(-u)]f(u)\nonumber\\
&+&i\int_{0}^{\infty}du[f(iu)-f(-iu)]\sum_{j=\pm1}^{}\frac{g(i j u)}{e^{2\pi(u+i j\tilde{\beta})}-1}  \  .
\label{Abel-Plana}
\end{eqnarray}
For this case, we can identify
\begin{eqnarray}
g(u)&=&e^{2\pi i u\Delta z/L} \nonumber\\
	f(u)&=&\frac{e^{-i\Delta t\sqrt{\lambda^2+p^2+\vec{k}^2+(2\pi u/L)^2}}}{\sqrt{\lambda^2+p^2+\vec{k}^2+(2\pi u/L)^2}} \  , 
\end{eqnarray}

Using \eqref{Abel-Plana}, we can write the Wightman function as
\begin{equation}
	W(x,x')=W_{cs}(x,x')+W_{c}(x,x') \ .
	\label{propagator}
\end{equation}
The first term represents the contribution due to the  AdS bulk without compactification, which, for our analysis, besides to present dependence on the magnetic fluxes also depends on the conical structure induced by the presence of the cosmic string. As to the second term it is induced by the compactification and contains contributions due to the magnetic flux enclosed by the compactified axis. Both expressions are explcitly wrtten in \eqref{propagator-cs} and \eqref{propagator-compactification}, respectively. 

The first term on the right hand of \eqref{propagator}, derived from the first integral of \eqref{Abel-Plana}, can be written as,
\begin{eqnarray}
W_{cs}(x,x')&=&\frac{q(ww')^{\frac{D}{2}}e^{ -ieA_z\Delta z}}{2(2\pi)^{D-2}a^{D-1}}\int d\vec{k}e^{i\vec{k}\cdot\Delta\vec{x}_{\parallel}}\int_0^{\infty}dppJ_{\nu}(pw)J_{\nu}(pw')
\sum_{n=-\infty}^{\infty}e^{inq\Delta\phi}\nonumber\\
&\times&\int_0^{\infty}d\lambda\lambda J_{q|n +\alpha|}(\lambda r)J_{q|n +\alpha|}(\lambda r')\int dk_z e^{ik_z\Delta z}\frac{e^{-i\Delta t\sqrt{\lambda^2+p^2+\vec{k}^2+k_z^2}}}{\sqrt{\lambda^2+p^2+\vec{k}^2+k_z^2}} \ ,
\label{propagator-cs}
\end{eqnarray}
where we have defined a new variable $k_z=2\pi u/L$.\footnote{For the case of vanishing magnetic fluxes and the absence of cosmic string, i.e, $q=1$, the expression \eqref{propagator-cs} reduces itself to the Wightmann function in a Ads bulk only.} Now performing a Wick rotation, and  using the following identity below,
\begin{equation}
\frac{e^{-\Delta\tau\omega}}{\omega}=\frac2{\sqrt{\pi}}\int_0^\infty ds e^{-s^2\omega^2-\Delta\tau^2/(4s^2)}  \  ,
	\label{identity}
\end{equation} 
the integration over $k_z$ can be evaluated, and the result is
\begin{eqnarray}
	W_{cs}(x,x')&=&\frac{q(ww')^{\frac{D}{2}}e^{ -ieA_z\Delta z}}{(2\pi)^{D-2}a^{D-1}}\int d\vec{k}e^{i\vec{k}\cdot\Delta\vec{x}_{\parallel}}\int_0^{\infty}dppJ_{\nu}(pw)J_{\nu}(pw')\sum_{n=-\infty}^{\infty}e^{inq\Delta\phi}\nonumber\\
	&\times&\int_0^{\infty}d\lambda\lambda J_{q|n +\alpha|}(\lambda r)J_{q|n +\alpha|}(\lambda r')\int_{0}^{\infty}\frac{ds}{s}e^{-s^2(\lambda^2+p^2+\vec{k}^2)-(\Delta z^2-\Delta t^2)/4s^2}\ .
\label{propagator-cs-2}
\end{eqnarray}

Now let us concentrate on the second term of \eqref{propagator}. Defining  again the variable  $k_z=2\pi u/L$, the integral over this variable must be considered in two different intervals: In the first interval $[0,\ \sqrt{\lambda^2+p^2+\vec{k}^2}]$, the integral vanishes, so it remains the contribution coming from the second interval, $[\sqrt{\lambda^2+p^2+\vec{k}^2}, \ \infty)$. So, taking into account this analysis we get,
\begin{eqnarray}
\label{W_compact}
	W_{c}(x,x')&=&\frac{q(ww')^{\frac{D}{2}}e^{ -ieA_z\Delta z}}{(2\pi)^{D-2}a^{D-1}}\int d\vec{k}e^{i\vec{k}\cdot\Delta\vec{x}_{\parallel}}\int_0^{\infty}dppJ_{\nu}(pw)J_{\nu}(pw')\nonumber\\
	&\times&\sum_{n=-\infty}^{\infty}e^{inq\Delta\phi}\int_0^{\infty}d\lambda\lambda J_{q|n +\alpha|}(\lambda r)J_{q|n +\alpha|}(\lambda r')\nonumber\\
	&\times&\int_{\sqrt{\lambda^2+p^2+\vec{k}^2}}^{\infty}dk_z\frac{\cosh{\big(\Delta t\sqrt{k_z^2-\lambda^2-p^2-\vec{k}^2}\big)}}{\sqrt{k_z^2-\lambda^2-p^2-\vec{k}^2}}
	\times\sum_{j=\pm 1}^{\infty}\frac{e^{-jk_z\Delta z}}{e^{Lk_z+2\pi ij\tilde{\beta}}-1}.
\end{eqnarray}
Developing the series expansion $(e^y-1)^{-1}=\sum_{l=1}^{\infty}e^{-ly}$, and with the help of \cite{Prud},  it is possible to integrate over $k_z$, obtaining
\begin{eqnarray}
W_{c}(x,x')&=&\frac{q(ww')^{\frac{D}{2}}e^{-ieA_z\Delta z}}{(2\pi)^{D-2}a^{D-1}}\int d\vec{k}e^{i\vec{k}\cdot\Delta\vec{x}_{\parallel}}\int_0^{\infty}dppJ_{\nu}(pw)J_{\nu}(pw')
\nonumber\\
&\times&\sum_{n=-\infty}^{\infty}e^{inq\Delta\phi}\int_0^{\infty}d\lambda\lambda J_{q|n +\alpha|}(\lambda r)J_{q|n +\alpha|}(\lambda r')\nonumber\\
&\times&\sum_{j=\pm 1}^{}\sum_{l=1}^{\infty}e^{-2\pi i\tilde{\beta} jl}K_{0}\big(\sqrt{(\lambda^2+p^2+\vec{k}^2)[(lL+j\Delta z)^2-\Delta t^2]}\big)  \  .
\label{propagator-compactification}
\end{eqnarray}
Clearly we notice that for $L\rightarrow\infty$, the function above vanishes. By using the integral representation below for the Macdonald function \cite{Grad},
\begin{equation}
K_{\nu}(x)=\frac{1}{2}\bigg(\frac{x}{2}\bigg)^{\nu}\int_{0}^{\infty}d\tau\frac{e^{-\tau-x^2/4\tau}}{\tau^{\nu+1}},
\label{representation-Macdonald}
\end{equation}
we can rewrite Eq.\eqref{propagator-compactification} as
\begin{eqnarray}
	W_{c}(x,x')&=&\frac{q(ww')^{\frac{D}{2}}e^{-ieA_z\Delta z}}{(2\pi)^{D-2}a^{D-1}}\int d\vec{k}e^{i\vec{k}\cdot\Delta\vec{x}_{\parallel}}\int_0^{\infty}dppJ_{\nu}(pw)J_{\nu}(pw')\nonumber\\
	&\times&\sum_{n=-\infty}^{\infty}e^{inq\Delta\phi}\int_0^{\infty}d\lambda\lambda J_{q|n +\alpha|}(\lambda r)J_{q|n +\alpha|}(\lambda r')\nonumber\\
&\times&	\sum_{j=\pm 1}^{}\sum_{l=1}^{\infty}e^{-2\pi i\tilde{\beta}jl}\int_{0}^{\infty}\frac{ds}{s}e^{-s^2(\lambda^2+p^2+\vec{k}^2)-[(lL+j\Delta z)^2-\Delta t^2]/4s^2}  \  .
	\label{propagator-compactification-2}
\end{eqnarray}

Substituting \eqref{propagator-cs-2} and \eqref{propagator-compactification-2} into \eqref{propagator}, and after some manipulations, we get a compact expression for the total Wigthman function given below,
\begin{eqnarray}
	W(x,x')&=&\frac{q(ww')^{\frac{D}{2}}e^{-ieA_z\Delta z}}{(2\pi)^{D-2}a^{D-1}}\int d\vec{k}e^{i\vec{k}\cdot\Delta\vec{x}_{\parallel}}\int_0^{\infty}dppJ_{\nu}(pw)J_{\nu}(pw')\nonumber\\
	&\times&	\sum_{n=-\infty}^{\infty}e^{inq\Delta\phi}\int_0^{\infty}d\lambda\lambda J_{q|n +\alpha|}(\lambda r)J_{q|n +\alpha|}(\lambda r')\nonumber\\
	&\times&\sum_{l=-\infty}^{\infty}e^{-2\pi i\tilde{\beta}l}\int_{0}^{\infty}\frac{ds}{s}e^{-s^2(\lambda^2+p^2+\vec{k}^2)-[(lL+j\Delta z)^2-\Delta t^2]/4s^2}   \  .
\end{eqnarray}
Now using the integral below \cite{Grad},
\begin{equation}
\int_0^{\infty}d\eta\eta e^{-\eta^2s^2}J_{\gamma}(\eta\rho)J_{\gamma}(\eta\rho') = \frac{e^{-\frac{(\rho^2 + \rho'^2)}{4s^2}}}{2s^2}I_{\gamma}\left(\frac{\rho\rho'}{2s^2}\right)  \  ,
\label{id2}
\end{equation}
we can integrate over $\lambda$, $p$ and $\vec{k}$, obtaining
\begin{eqnarray}
	W(x,x')&=&\frac{qe^{-ieA_z\Delta z}}{2(2\pi)^{\frac{D}{2}}a^{D-1}}\bigg(\frac{ww'}{rr'}\bigg)^{\frac{D}{2}}\sum_{l=-\infty}^{\infty}e^{-2\pi i\tilde{\beta}l}\int_{0}^{\infty}d\chi \chi^{\frac{D}{2}-1}e^{-\chi u_{l}^{2}/2rr'}
	I_{\nu}\bigg(\frac{ww'}{rr'}\chi\bigg)\nonumber\\
	&\times&\sum_{n=-\infty}^{\infty}e^{iqn\Delta\phi}I_{q|n+\alpha|}(\chi)  \  , 
	\label{propagator-to-sum}
\end{eqnarray}
where we have introduced a new variable $\chi=rr'/2s^2$, and defined
\begin{equation}
	u_{l}^{2}=r^2+r'^2+w^2+w'^2+(lL+\Delta z)^2+\Delta \vec{x}^{2}_{\parallel}-\Delta t^2  \ .
\end{equation}
The parameter $\alpha$ in Eq.\eqref{const} can be written in the form
\begin{equation}
\alpha=n_{0}+\varepsilon, \ \textrm{with}\ |\varepsilon|<\frac{1}{2},
\label{const-2}
\end{equation}
being $n_{0}$ is an integer number. This allow us to sum over the quantum number $n$ in \eqref{propagator-to-sum}, using the result obtained in \cite{deMello:2014ksa}, given below,
\begin{eqnarray}
&&\sum_{n=-\infty}^{\infty}e^{iqn\Delta\phi}I_{q|n+\alpha|}(\chi)=\frac{1}{q}\sum_{k}e^{\chi\cos(2\pi k/q-\Delta\phi)}e^{i\alpha(2\pi k -q\Delta\phi)}\nonumber\\
	&-&\frac{e^{-iqn_{0}\Delta\phi}}{2\pi i}\sum_{j=\pm1}je^{ji\pi q|\varepsilon|}
	\int_{0}^{\infty}dy\frac{\cosh{[qy(1-|\varepsilon|)]}-\cosh{(|\varepsilon| qy)e^{-iq(\Delta\phi+j\pi)}}}{e^{\chi\cosh{(y)}}\big[\cosh{(qy)}-\cos{(q(\Delta\phi+j\pi))}\big]},
	\label{summation-formula}
\end{eqnarray}
where
\begin{equation}
	-\frac{q}{2}+\frac{\Delta\phi}{\Phi_{0}}\le k\le \frac{q}{2}+\frac{\Delta\phi}{\Phi_{0}}  \   .
\end{equation}
In short, the obtainment of the above expression is through the integral representation for the modified Bessel function \cite{Abra},
\begin{equation}
I_{q|n+\alpha|}(z)=\frac{1}{\pi }\int_{0}^{\pi }dy\;\cos (q|n+\alpha|
y)e^{z\cos y}-\frac{\sin (\pi q|n+\alpha|)}{\pi }\int_{0}^{\infty
}dye^{-z\cosh y-q|n+\alpha| y}  \  ,  
\end{equation}
following by the summation of the quantum number $n$ and some additional intermediate steps.

Thus, the substitution of \eqref{summation-formula} into \eqref{propagator-to-sum}, allow us to integrate over $\chi$ with the help of \cite{Grad}, yielding
\begin{eqnarray}
	W(x,x')&=&\frac{e^{-ieA_z\Delta z}}{(2\pi)^{\frac{D+1}{2}}a^{D-1}}\sum_{l=-\infty}^{\infty}e^{-2\pi i\tilde{\beta}l}\Bigg\{\sum_{k}e^{i\alpha(2\pi k-q\Delta\phi)}F_{\nu-1/2}^{(D-1)/2}({u}_{lk})\nonumber\\
	&-&q\frac{e^{-iqn_{0}\Delta\phi}}{2\pi i}\sum_{j=\pm1}je^{ji\pi q|\varepsilon|}
	\int_{0}^{\infty}dy\frac{\cosh{[(1-|\varepsilon|)qy]}-\cosh{(|\varepsilon|q y)e^{-iq(\Delta\phi+j\pi)}}}{\cosh{(qy)}-\cos{(q(\Delta\phi+j\pi))}}\nonumber\\
	&\times&F_{\nu-1/2}^{(D-1)/2}({u}_{ly})\Bigg\},
	\label{full-propagator}
\end{eqnarray}
where we have introduced the notation
\begin{eqnarray}
F_{\gamma}^{\mu}(u)&=&e^{-i\pi\mu}\frac{Q^{\mu}_{\gamma}(u)}{(u^2-1)^{\mu/2}}  \nonumber\\  
&=&\frac{\sqrt{\pi}\Gamma(\gamma+\mu+1)}{2^{\gamma+1}\Gamma(\gamma+3/2)u^{\gamma+\mu+1}}F\bigg(\frac{\gamma+\mu}{2}+1,\frac{\gamma+\mu+1}{2};\gamma+\frac{3}{2};\frac{1}{u^{2}}\bigg).
\label{function-2}
\end{eqnarray}
being $Q_{\gamma}^{\mu}(u)$ the associated Legendre function of second kind and $F(a,b;c;z)$ the hypergeometric function \cite{Abra}. In \eqref{full-propagator}, the arguments of the function $F_{\gamma}^{\mu}$ are given by
\begin{eqnarray}
u_{lk}&=&1+\frac{r^2+r'^2-2rr'\cos{(2\pi k/q-\Delta\phi)}+\Delta w^2+(lL+\Delta z)^2+\Delta\vec{x}^{2}_{\parallel}-\Delta t^2}{2ww'}\nonumber\\
u_{ly}&=&1+\frac{r^2+r'^2+2rr'\cosh{(y)}+\Delta w^2+(lL+\Delta z)^2+\Delta\vec{x}^{2}_{\parallel}-\Delta t^2}{2ww'}.
\end{eqnarray}
So, \eqref{full-propagator} is the the most compact expression to the Wightman function. In this format the $l=0$ component corresponds to the contribution due to the cosmic string only, and $l\neq0$ is the contribution due to the compactification.

Having obtained the above result, we are in position to calculate the induced current densities. This new subject is left to the next sections.

\section{Bosonic Current}
\label{sec3}
The bosonic current density operator is given by,
\begin{eqnarray}
\hat{j_{\mu }}(x)&=&ie\left[\hat{\varphi} ^{*}(x)D_{\mu }\hat{\varphi} (x)-
(D_{\mu }\hat{\varphi})^{*}\hat{\varphi}(x)\right] \nonumber\\
&=&ie\left[\hat{\varphi}^{*}(x)\partial_{\mu }\hat{\varphi} (x)-\hat{\varphi}(x)
(\partial_{\mu }\hat{\varphi}(x))^{*}\right]-2e^2A_\mu(x)|\hat{\varphi}(x)|^2 \   .
\label{eq20}
\end{eqnarray}
Its vacuum expectation value (VEV) can be evaluated in terms of the positive frequency Wightman function as exhibited below:
\begin{equation}
\left\langle j_{\mu}(x) \right\rangle=ie\lim_{x'\rightarrow x}
\left\{(\partial_{\mu}-\partial_{\mu}')W(x,x')+2ieA_\mu W(x,x')\right\} \ .
\label{eq21}
\end{equation}

As we will see, this VEV is a periodic function of the magnetic fluxes $\Phi_\phi$ and
$\Phi_z$ with period equal to the quantum flux. This can be observed writing 
the parameter $\alpha$ as in \eqref{const-2}.

\subsection{Charge Density}
Let us begin with the calculation of the charge density. Since $A_{0}=0$, we have
\begin{equation}
\langle j_{0}(x)\rangle=ie\lim\limits_{x'\rightarrow x}(\partial_{t}-\partial'_{t})W(x,x').
\label{charge-density}
\end{equation}
Substituting Eq.\eqref{full-propagator} into the above expression, taking the time derivatives and finally the coincidence limit, we obtain
a divergent result.  To avoid this problem a regularization procedure is necessary. Many regularization procedure can be applied; however for the present problem the most convenient is the Pauli-Villars gauge-invariant (PV) one. Adopting this procedure, regulator fields with large masses are introduced. The number of these fields depends on the specific problem. As we will see below, a single regulator field with mass $M$ is sufficient. By using PV the regularized VEV of the charge density reads, 
\begin{eqnarray}
\langle j^{0}(x)\rangle_{Reg}&=&\frac{2ie}{(2\pi)^{\frac{D+1}{2}}a^{D+1}} \lim\limits_{t'\rightarrow t}\Delta t\sum_{l=-\infty}^{\infty}e^{-2\pi i\tilde{\beta}l}\left[\sum_{k}e^{2\pi ki\alpha}\sum_{n=0,1}c_{n}F_{\nu_{(n)}-1/2}^{(D+1)/2}(\tilde{u}_{lk})\right.\nonumber\\
&-&\left.\frac{q}{2\pi i}\sum_{j=\pm1}je^{ji\pi q|\varepsilon|}
\int_{0}^{\infty}dy\frac{\cosh{[(1-|\varepsilon|)qy]}-\cosh{(|\varepsilon|qy)e^{-iqj\pi}}}{\cosh{(qy)}-\cos{(qj\pi)}}\right.\nonumber\\
&\times&\left.\sum_{n=0,1}c_nF_{\nu_{(n)}-1/2}^{(D+1)/2}(\tilde{u}_{ly})\right]   ,
\label{charge-density-2}
\end{eqnarray}
where $c_0=1$, $\nu_{(0)}=\nu$, given by \eqref{const}, $c_{1}=-1$ and $\nu_{(1)}$ is the corresponding parameter associated with the mass $M$. Moreover, in \eqref{charge-density-2}
the arguments of the functions are,
\begin{eqnarray}
\label{variables-u}
\tilde{u}_{lk}&=&1+\frac{4r^2\sin^2{(\pi k/q)}+(lL)^2-\Delta t^2}{2w^2}\nonumber\\
\tilde{u}_{ly}&=&1+\frac{4r^2\cosh^2{(y)}+(lL)^2-\Delta t^2}{2w^2}  \ .
\end{eqnarray}
In the obtainment of the above result we have used the following relation
\begin{equation}
\partial_{x}F_{\gamma}^{\mu}(u(x))=-(\partial_{x}u(x))F_{\gamma}^{\mu+1}(u(x)) \  ,
\label{derivative}
\end{equation}
by using the recurrence relations for the associated Legendre function of second kind \cite{Abra}.

We can see from  \eqref{variables-u}, that the arguments of the functions $F_{\nu_{(n)}-1/2}^{(D+1)/2}$ above, for $l\neq0$, are bigger than unity. Consequently the corresponding compactified contributions inside the brackets of \eqref{charge-density-2} are finite, providing a vanishing contribution for the charge density when we take the time coincidence limit, $\Delta t\to 0$. On the other hand for cosmic string contribution ($l=0$) for $k=0$ and for $k\neq 0$ but with $r=0$, the arguments of the functions go to unit for the time coincidence limit. However, in the limit of argument close to $1$, by using the asymptotic formula for the hypergeometric function, we get a divergent result below which does not depend on the parameter $\nu$, 
\begin{eqnarray}
F_{\nu-1/2}^{(D+1)/2}(u)\approx\frac{\Gamma((D+1)/2)}{2(u-1)^{(D+1)/2}} \  .
\end{eqnarray} 
So, the divergent behavior of the combination, $F_{\nu_{(0)}-1/2}^{(D+1)/2}(u)- F_{\nu_1-1/2}^{(D+1)/2}(u)$, is canceled. Finally taking the time coincidence limit in \eqref{charge-density-2} these contributions also provide a vanishing result. So, we conclude that the charge density vanishes.

Following similar procedure we also can prove that no radial current density, $\langle j^r\rangle$, currents densities along $w$, $\langle j^w\rangle$, and extra dimensions, $\langle j^i\rangle$ for $i=5, \ 6, \ ... \ $, are induced by this system. 

\subsection{Azimuthal Current}
The VEV of the azimuthal current density is given by
\begin{equation}
\langle j_{\phi}(x)\rangle =ie\lim\limits_{x'\rightarrow x}\{(\partial_{\phi}-\partial'_{\phi	})W(x,x')+2ieA_{\phi}W(x,x')\}  \  .
\label{eqn:azimuthal-current}
\end{equation}

Substituting \eqref{wight2} into the above equation, we formally can express the azimuthal current as,
\begin{eqnarray}
	\langle j_{\phi}(x)\rangle&=&-\frac{qea^{1-D}w^{D}}{(2\pi)^{D-3}L} \sum_{n=-\infty}^{\infty}q(n+\alpha)\int d\vec{k}\int_{0}^{\infty}\lambda J_{q|n+\alpha|}^{2}(\lambda r)d\lambda
	\nonumber\\
	&\times&\int_{0}^{\infty}pJ_{\nu}^{2}(pw)dp
	\sum_{l=-\infty}^{\infty}\frac{1}{\sqrt{\lambda^{2}+p^{2}+\vec{k}^{2}+\tilde{k}_{l}^{2}}}  \  .
	\label{eqn:azimuthal-current-2}
\end{eqnarray}

Identifying $g(u)=1$ and
\begin{equation}
	f(u)=\frac{1}{\sqrt{\lambda^{2}+p^{2}+\vec{k}^{2}+(2\pi u/L)^{2}}},
	\label{function}
\end{equation}
we can use \eqref{Abel-Plana} to develop the summation on the quantum number $l$. Doing this, the VEV is decomposed  as
\begin{equation}
	\langle j_{\phi}(x)\rangle=	\langle j_{\phi}(x)\rangle_{cs}+\langle j_{\phi}(x)\rangle_{c}  \   ,
	\label{current-decomposition}
\end{equation}
where $\langle j_{\phi}(x)\rangle_{cs}$ corresponds the contribution from the cosmic string without compactification, which comes from the first integral on the right hand side of the Eq.\eqref{Abel-Plana}. This component reads,
\begin{eqnarray}
	\langle j_{\phi}(x)\rangle_{ cs}&=&-\frac{2qew^{D}}{(2\pi)^{D-2}a^{D-1}} \sum_{n=-\infty}^{\infty}q(n+\alpha)\int d\vec{k}\int_{0}^{\infty}d\lambda\lambda J_{q|n+\alpha|}^{2}(\lambda r)
	\nonumber\\
	&\times&\int_{0}^{\infty}pJ_{\nu}^{2}(pw)dp\int_{0}^{\infty}\frac{dk_z}{\sqrt{\lambda^{2}+p^{2}+\vec{k}^{2}+k_z^{2}}} \   ,
	\label{eqn:azimuthal-current-first-part}
\end{eqnarray}
where we have defined $k_z=2\pi u/L$. 

Using the identity below, 
\begin{equation}
\frac{1}{\sqrt{\lambda^{2}+p^{2}+\vec{k}^{2}+k_z^{2}}
	} = \frac{2}{\sqrt{\pi}}\int_{0}^{\infty}dse^{-s^2(\lambda^2+p^2+\vec{k}^2+k_z^2)},
\label{eqn:identity}
\end{equation}
and \eqref{id2} we can perform the integrations over all but $s$ variable, obtaining
\begin{eqnarray}
\langle j_{\phi}(x)\rangle_{cs}=-\frac{eq^2w^D}{(2\pi)^{\frac{D}{2}}a^{D-1}r^D}\int_{0}^{\infty}d\chi \chi^{\frac{D}{2}-1}e^{-\chi[1+(w/r)^2]}I_{\nu}\bigg(\frac{w^2\chi}{r^2}\bigg)\sum_{n=-\infty}^{\infty}(n+\varepsilon)I_{q|n+\varepsilon|}(\chi)  \  .
\label{azimuthal-current-first-part-2}
\end{eqnarray}
We have written $\alpha$ in the form \eqref{const-2} and also introduced a new variable, $\chi=r^2/2s^2$. In \cite{Braganca:2014qma} it has been derived a compact expression for the summation over the quantum number $n$. This result we reproduce below,
\begin{eqnarray}
\sum_{n=-\infty}^{\infty}(n+\varepsilon)I_{q|n+\varepsilon|}(\chi)&=&\frac{2\chi}{q^2}\sideset{}{'}\sum_{j=1}^{[q/2]}\sin{(2\pi j/q)}\sin{(2\pi j\varepsilon)}e^{\chi\cos(2\pi j/q)}\nonumber\\
&+&\frac{\chi}{q\pi}\int_{0}^{\infty}dy\sinh{(y)}\frac{e^{-\chi\cosh{(y)}}g(q,\varepsilon,y)}{\cosh{(qy)}-\cos{(\pi q)}}  \  ,
\label{Summation-formula}
\end{eqnarray}
where $[q/2]$ represents the integer part of $q/2$, and the prime on the sign of
the summation means that in the case $q=2p$ the term $k=q/2$ should be
taken with the coefficient $1/2$. Moreover the function, $g(q,\varepsilon,y)$, is defined as
\begin{equation}
	g(q,\varepsilon,y)=\sin{(q\pi  \varepsilon)} \sinh{((1-|\varepsilon|)qy)}-\sinh{(q\varepsilon y)}\sin{((1-|\varepsilon|)\pi q)}.
	\label{eqn:summation-formula-2}
\end{equation}

\begin{figure}[!htb]
	\begin{center}
		\centering
		\includegraphics[scale=0.40]{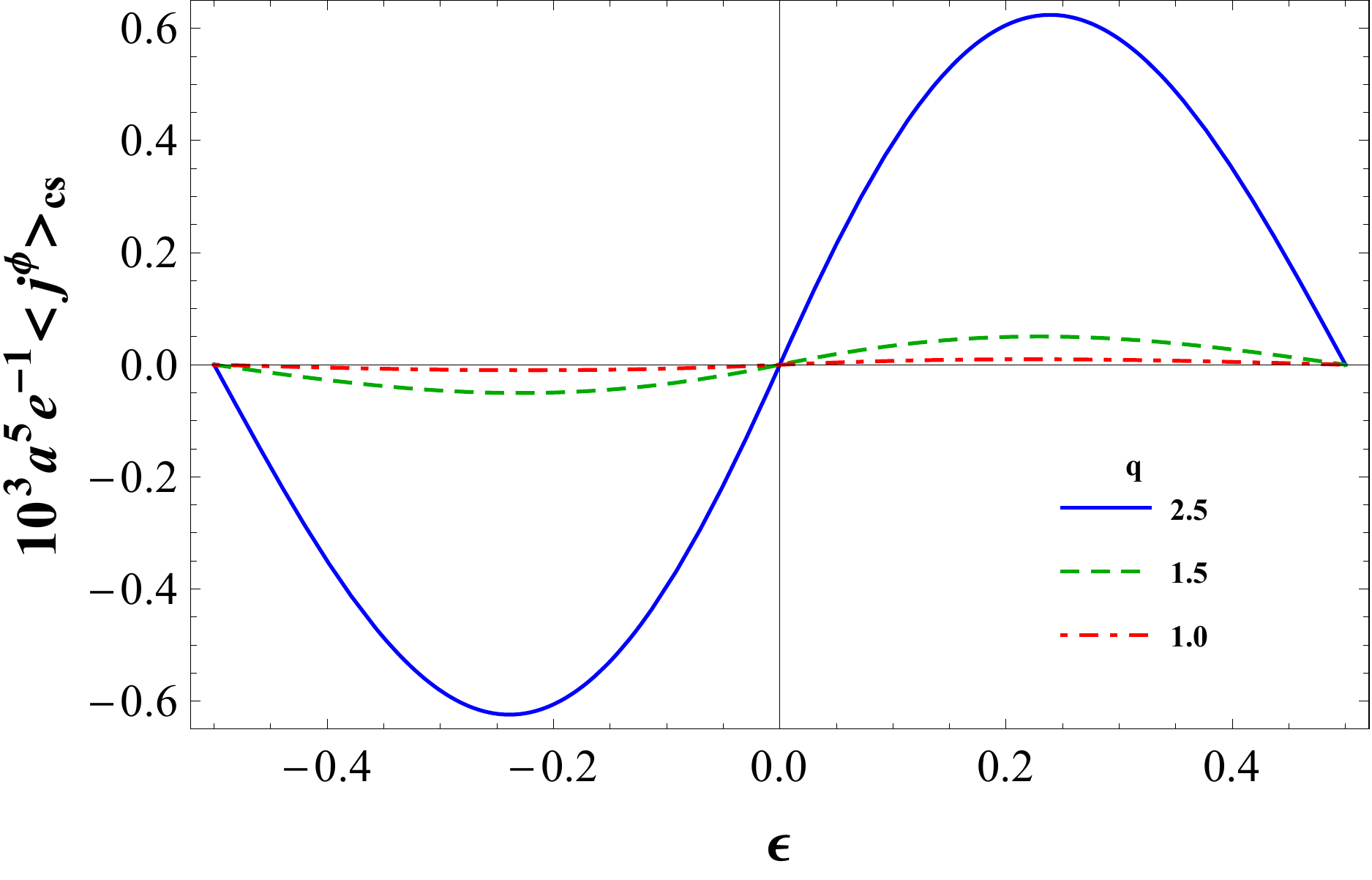}
		\caption{The azimuthal current density without compactification for $D=4$ in Eq.\eqref{azimuthal-current-first-part-3} is plotted, in units of ``$ea^{-5}$'', in terms of $\varepsilon$,  for $r/w=1$, $ma=1$, $\xi=0$ and $q=1,1.5$ and $2.5$.}
		\label{fig1}
	\end{center}
\end{figure}

\begin{figure}[!htb]
	\begin{center}
	\includegraphics[scale=0.35]{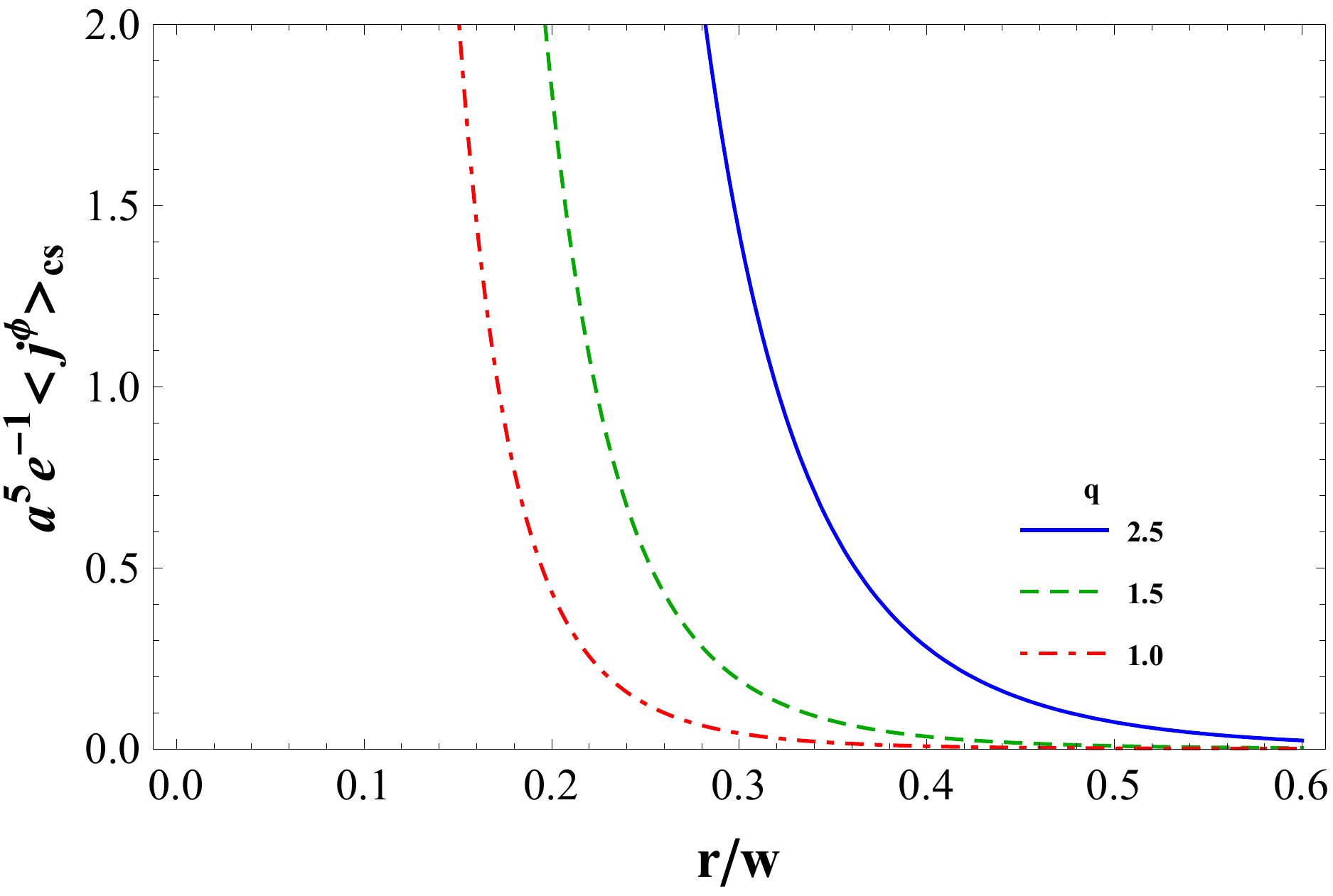}
	\quad
	\includegraphics[scale=0.35]{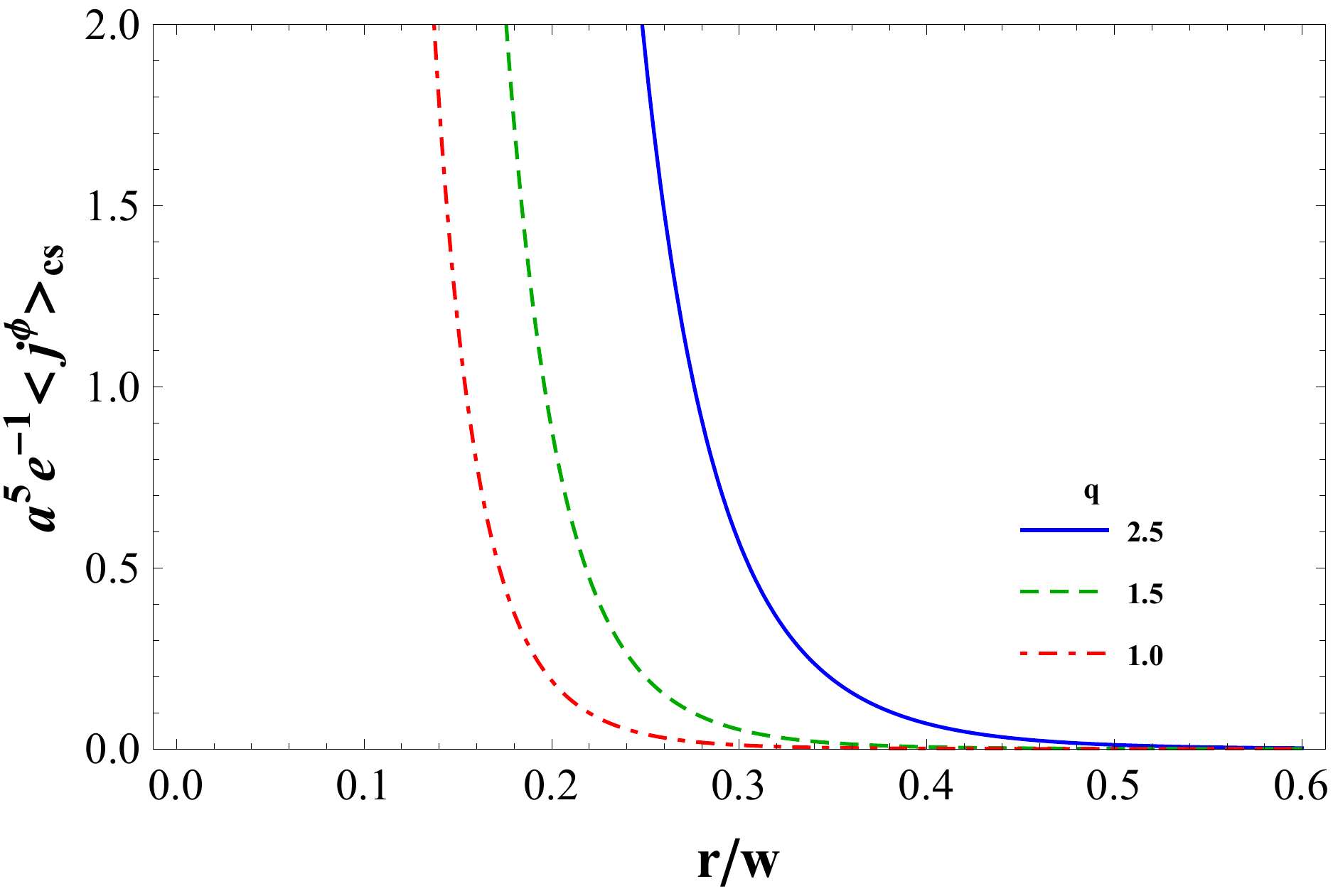}
	\caption{The azimuthal current density without compactification for $D=4$ in Eq.\eqref{azimuthal-current-first-part-3} is plotted, in units of ``$ea^{-5}$'', in terms of the proper distance, $r/w$, for $\varepsilon=0.25$, $\xi=0$ and $q=1,1.5,2.5.$ The plot on the left is for $ma=1$ while the plot on the right is for $ma=5$.}
	\label{fig2}
	\end{center}
\end{figure}

Substituting the above result into \eqref{azimuthal-current-first-part-2} and with the help of \cite{Grad}, we get,
\begin{eqnarray}
	\langle j^{\phi}(x)\rangle_{cs}&=&\frac{4ea^{-(1+D)}}{(2\pi)^{\frac{D+1}{2}}}\Bigg[\sideset{}{'}\sum_{j=1}^{[q/2]}\sin{(2\pi j/q)}\sin{(2\pi j\varepsilon)}F_{\nu-1/2}^{(D+1)/2}(u_{j})\nonumber\\
	&+&\frac{q}{\pi}\int_{0}^{\infty}dy\frac{\sinh{(2y)}g(q,\varepsilon,2y)}{\cosh{(2qy)}-\cos{(\pi q)}}F_{\nu-1/2}^{(D+1)/2}(u_{y})\Bigg],
	\label{azimuthal-current-first-part-3}
\end{eqnarray}
where the arguments of the functions are given by
\begin{eqnarray}
	u_{j}&=&1+2(r/w)^2\sin^2{(\pi j/q)}, \nonumber\\
	u_{y}&=&1+2(r/w)^2\cosh^2{(y)}  \  .
\end{eqnarray}
From \eqref{azimuthal-current-first-part-3}, we can see that $\langle j^{\phi}(x)\rangle_{cs}$ is an odd function of $\varepsilon$ with period equal to quantum flux, $\Phi_0=2\pi/e$; moreover,  for $1\leq q<2$ the first term on the right hand side is absent. In Fig. \ref{fig1} we exhibit the behavior of the azimuthal current as function of $\varepsilon$, considering $D=4$, the minimal curvature coupling, $\xi=0$, $r/w=ma=1$ and for different values of $q$. As we can see the intensity of the current depends strongly on the value of $q$. Increasing $q$ its intensity also increases. 
\\
In Fig.\ref{fig2} we plot the behavior the azimuthal current, $\langle j^{\phi}(x)\rangle_{cs}$ as function of $r/w$ for $D=4$, considering $\varepsilon=0.25$ and $\xi=0$, for different values of the parameter $q$. In the left panel we adopted $ma=1$, and in the right panel $ma=5$.

After the numerical analyses for $\langle j^{\phi}(x)\rangle_{cs}$, we will develop its behavior for some specific regimes of the physical variables. We start considering $r/w\rightarrow0$. We can use the asymptotic formula for the hypergeometric function for small arguments \cite{Abra} to rewrite Eq.\eqref{azimuthal-current-first-part-3} as
\begin{eqnarray}
\langle j^{\phi}(x)\rangle_{cs}&\approx& \frac{4e\Gamma(\frac{D+1}{2})}{(4\pi)^{\frac{D+1}{2}}}\bigg(\frac{w}{ar}\bigg)^{D+1}\Bigg[\sideset{}{'}\sum_{j=1}^{[q/2]}\frac{\cot{(\pi j/q)}\sin{(2\pi j\varepsilon)}}{\sin^{D-1}{(\pi j/q)}}\nonumber\\
&+&\frac{q}{\pi}\int_{0}^{\infty}dy\frac{\tanh{(y)}}{\cosh^{D-1}{(y)}}\frac{g(q,\varepsilon,2y)}{\cosh{(2qy)}-\cos{(\pi q)}}\Bigg]  \  .
\label{azimuthal-current-first-asymptotic}
\end{eqnarray}
Apart form the conformal factor, $(w/a)^{(D+1)}$, the above expression coincides with the corresponding one in Minkowski background for points near the string \cite{Braganca:2014qma}. On the other hand, for $r/w\gg1$, we have
\begin{eqnarray}
\langle j^{\phi}(x)\rangle_{cs}&\approx& \frac{2^{1-2\nu}e\Gamma(D/2+\nu+1)}{(4\pi)^{\frac{D}{2}}\Gamma(\nu+1)
	a^{D+1}}\bigg(\frac{w}{r}\bigg)^{D+2\nu+2}\bigg\{\sideset{}{'}\sum_{j=1}^{[q/2]}\frac{\cot{(\pi j/q)}\sin{(2\pi j\varepsilon)}}{\sin^{D+2\nu}{(\pi j/q)}}\nonumber\\
&+&\frac{q}{\pi}\int_{0}^{\infty}dy\frac{\tanh{(y)}}{\cosh^{D+2\nu}(y)}\frac{g(q,\varepsilon,2y)}{\cosh{(2qy)}-\cos{(\pi q)}}\Bigg\}  \  .
\label{azimuthal-current-first-asymptotic-2}
\end{eqnarray}
Another interesting asymptotic behavior is for $\nu\gg1$. For this case \eqref{azimuthal-current-first-part-3} reads,
\begin{eqnarray}
\langle j^{\phi}(x)\rangle_{cs}&\approx& \frac{2e\nu^{D/2}}{(2\pi)^{\frac{D}{2}}a^{1+D}}\Bigg[\sideset{}{'}\sum_{j=1}^{[q/2]}\sin{(2\pi j/q)}\sin{(2\pi j\varepsilon)}\frac{\big(u_{j}+\sqrt{u_{j}^{2}-1}\big)^{-\nu}}{(u_{j}^{2}-1)^{\frac{D+2}{4}}}\nonumber\\
&+&\frac{q}{\pi}\int_{0}^{\infty}dy\frac{\sinh{(2y)}g(q,\varepsilon,2y)}{\cosh{(2qy)}-\cos{(\pi q)}}\frac{\big(u_{y}+\sqrt{u_{y}^{2}-1} \big)^{-\nu}}{(u_{y}^{2}-1)^{\frac{D+2}{4}}}\Bigg]  \  .
\label{azimuthal-current-first-part-4}
\end{eqnarray}

For a conformally coupled massless scalar field we have $\nu=1/2$, and by expressing the associated Legendre function in terms of  hypergeometric function \cite{Abra,Grad}, we can  write a more convenient expression for $F_{\nu-1/2}^{(D+1)/2}(u)$ \cite{deMello:2014hya}, given by
\begin{eqnarray}
	F_{0}^{(D+1)/2}(u)=-\frac{\Gamma\big(\frac{D+1}{2}\big)}{2}\bigg[(1+u)^{-(D+1)/2}-(u-1)^{-(D+1)/2}\bigg].
	\label{function-3}
\end{eqnarray} 
Substituting \eqref{function-3} into \eqref{azimuthal-current-first-part-3}, we obtain
\begin{eqnarray}
\label{j_conf}
	\langle j^{\phi}(x)\rangle_{cs}&=&\bigg(\frac{w}{a}\bigg)^{D+1} \Bigg\{\frac{4e\Gamma\big(\frac{D+1}{2}\big)}{(4\pi)^{\frac{D+1}2}{r^{(D+1)}}}\Bigg[\sideset{}{'}\sum_{j=1}^{[q/2]}\frac{\cos{(\pi j/q)}\sin(2\pi j\varepsilon)}{\sin^{D}(\pi j/q)}
	\nonumber\\
	&+&\frac{q}{\pi}\int_{0}^{\infty}dy\frac{\sinh{(y)}}{\cosh{(2qy)-\cos{(\pi q)}}}\frac{g(q,\varepsilon,2y)}{\cosh^{D}{(y)}}\Bigg]
	\nonumber\\
&-&\frac{2e\Gamma\big(\frac{D+1}{2}\big)}{(4\pi)^{\frac{D+1}2}{r^{(D+1)}}}\Bigg[\sideset{}{'}\sum_{j=1}^{[q/2]}\sin{(2\pi j/q)\sin{(2\pi j\varepsilon)}\bigg(\frac{w^2}{r^2}+\sin^2(\pi j/q)\bigg)^{-\frac{D+1}{2}}}\nonumber\\
&	&+\frac{q}{\pi}\int_{0}^{\infty}dy\frac{\sinh{(2y)}g(q,\varepsilon,2y)}{\cosh{(2qy)-\cos{(\pi q)}}}\bigg(\frac{w^2}{r^2}+\cosh^2(y)\bigg)^{-\frac{D+1}{2}}\Bigg]
	\Bigg\}  \  .
\end{eqnarray}
We notice that two different set of contributions appear in the expression above. Apart from the conformal factor, the first set coincides with the induced massless scalar azimuthal current in the Minkowski background \cite{Braganca:2014qma}. It is divergent for $r\to 0$. As to the second set, it is a new contribution. This part is induced by the boundary located at $w=0$. It is finite at the string's core for $w\neq 0$. In addition, for $r\gg w$ this part tends to cancel the first one.  Finally, taking $D=4$ in the above expression, we obtain
 \begin{eqnarray}
 	\langle j^{\phi}(x)\rangle_{cs}&=&\bigg(\frac{w}{a}\bigg)^{5} \Bigg\{\frac{3e}{32\pi^{2}r^5}\Bigg[\sideset{}{'}\sum_{j=1}^{[q/2]}\frac{\cot{(\pi j/q)}\sin(2\pi j\varepsilon)}{\sin^{3}(\pi j/q)}
 	\nonumber\\
 	&+&\frac{q}{\pi}\int_{0}^{\infty}dz\frac{\tanh{(z)}}{\cosh{(2qz)-\cos{(\pi q)}}}\frac{g(q,\epsilon,2z)}{\cosh^{3}{(z)}}\Bigg]
 	\nonumber\\
 &	-&\frac{3e}{64\pi^{2}r^5}\Bigg[\sideset{}{'}\sum_{j=1}^{[q/2]}\sin{(2\pi j/q)\sin{(2\pi j\varepsilon)}\bigg(\frac{w^2}{r^2}+\sin^2(\pi j/q)\bigg)^{-5/2}}\nonumber\\
 	&+&\frac{q}{\pi}\int_{0}^{\infty}dy\frac{\sinh{(2y)}g(q,\varepsilon,2y)}{\cosh{(2qy)-\cos{(\pi q)}}}\bigg(\frac{w^2}{r^2}+\cosh^2(y)\bigg)^{-5/2}\Bigg]\Bigg\} \   .
\end{eqnarray}

The compactified contribution for the azimuthal current, $\langle j_{\phi}(x)\rangle_{c}$, can be obtained using \eqref{W_compact}. So, we have,
\begin{eqnarray}
	\langle j_{\phi}(x)\rangle_{c}&=&-\frac{2qew^{D}} {(2\pi)^{D-2}a^{D-1}}\sum_{n=-\infty}^{\infty}q(n+\alpha)\int d\vec{k}\int_{0}^{\infty}\lambda J^{2}_{q|n+\alpha|}(\lambda r)d\lambda\nonumber\\
	&\times&\int_{0}^{\infty}p J^{2}_{\nu}(pw)dp\int_{\sqrt{\lambda^2+p^2+\vec{k}^2}}^{\infty}\frac{dk_z}{\sqrt{k_z^2-\lambda^2-p^2-\vec{k}^2}}\sum_{j=\pm1}^{}\frac{1}{e^{Lk_z+2\pi ij\tilde{\beta}}-1}.
	\label{compactification-contribuition}
\end{eqnarray}

To proceed with our analysis, it is necessary to use the series expansion $(e^y-1)^{-1}=\sum_{l=1}^{\infty}e^{-ly}$, and with the help of \cite{Grad} we can integrate over $k_z$, obtaining
\begin{eqnarray}
\langle j_{\phi}(x)\rangle_{c}&=&- \frac{4qew^{D}}{(2\pi)^{D-2}a^{D-1}}\sum_{l=1}^{\infty}\cos(2\pi l\tilde{\beta})\sum_{n=-\infty}^{\infty}q(n+\alpha)\int d\vec{k}\int_{0}^{\infty} d\lambda\lambda J^{2}_{q|n+\alpha|}(\lambda r)\nonumber\\
&\times&\int_{0}^{\infty}dp \ p J^{2}_{\nu}(pw) \ K_{0}\left(lL\sqrt{\lambda^2+p^2+\vec{k}^2}\right)  \  .
\label{compactification-contribuition-2}
\end{eqnarray}
Using the integral representation for the Macdonald function given in \eqref{representation-Macdonald}, it is possible to integrate over the variables $ \lambda$, $p$ and  $\vec{k}$, getting
\begin{eqnarray}
\langle j_{\phi}(x)\rangle_{c}&=&- \frac{2q^2ew^D}{(2\pi)^{D/2}a^{D-1}r^D} \sum_{l=1}^{\infty}\cos(2\pi l\tilde{\beta})\int_{0}^{\infty} d\chi\chi^{(D-2)/2}e^{-\chi[1+(l^2L^2+2w^2)/2r^2]}\nonumber\\
&\times&I_{\nu}\bigg(\frac{w^2\chi}{r^2}\bigg)\sum_{n=-\infty}^{\infty}(n+\varepsilon)I_{q|n+\varepsilon|}(\chi)  \  ,
\end{eqnarray}
where we have written $\alpha$ in the form of \eqref{const-2} and defined the variable, $\chi=\frac{2tr^2}{(lL)^2}$. Now using the sum formula given in \eqref{Summation-formula}, we are able to integrate over $\chi$, obtaining the final form of contribution to the azimuthal current density induced by the compactification: 
\begin{eqnarray}
	\langle j^{\phi}(x)\rangle_{c}&=& \frac{8ea^{-(D+1)}}{(2\pi)^{(D+1)/2}}\sum_{l=1}^{\infty}\cos(2\pi l\tilde{\beta})\Bigg[\sideset{}{'}\sum_{j=1}^{[q/2]}\sin(2\pi j/q)\sin(2\pi j\varepsilon)F_{\nu-1/2}^{(D+1)/2}(v_{lj})\nonumber\\
	&+&\frac{q}{\pi}\int_{0}^{\infty}dy\frac{\sinh{(2y)}g(q,\varepsilon,2y)}{\cosh(2qy)-\cos(\pi q)}F_{\nu-1/2}^{(D+1)/2}(v_{ly})\Bigg],
	\label{azimuthal-current-second-part}
\end{eqnarray}
being
\begin{eqnarray}
v_{lj}&=&1+\frac{(lL)^2+4r^2\sin^2{(\pi j/q)}}{2w^2}\nonumber\\
v_{ly}&=&1+\frac{(lL)^2+4r^2\cosh^2{(y)}}{2w^2}  \  .
\end{eqnarray}

\begin{figure}[!htb]
	\begin{center}
		\includegraphics[scale=0.40]{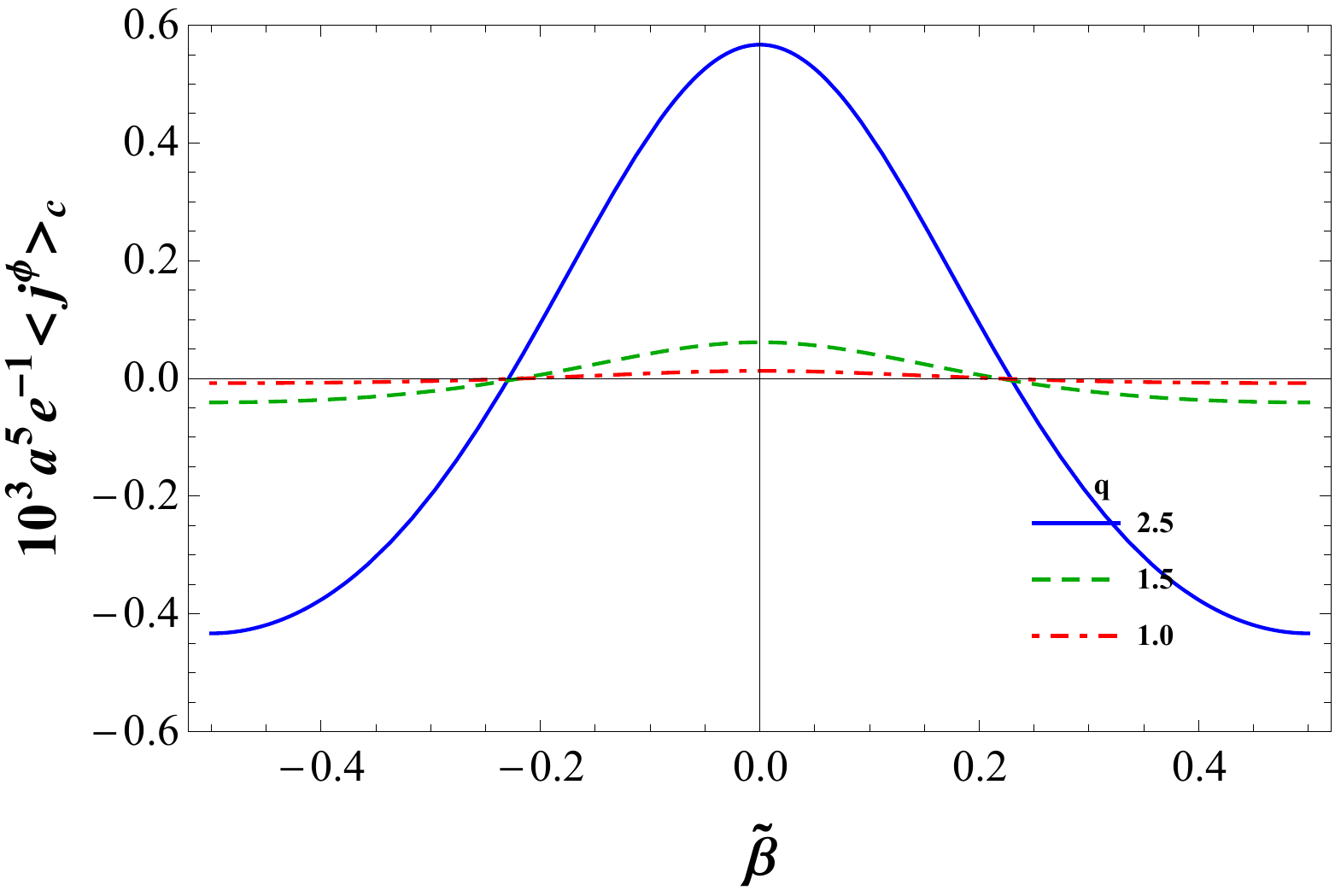}
	\caption{
		The azimuthal current density induced by compactification for $D=4$ is plotted, in units of ``$ea^{-5}$'', in terms of $\tilde{\beta}$ for $ma=1$, $\xi=0$, $\varepsilon=0.25$ and $q=1,1.5$ and $2.5$.}
	\label{fig3}
	\end{center}
\end{figure}

From the above expression we can see that the contribution due to the compactification on the azimuthal current density is an even function of the parameter $\tilde{\beta}$ and is an odd function of the magnetic flux along the core of the string, with period equal to $\Phi_0$. In particular, in the case of an untwisted bosonic field, $\langle j^{\phi }(x)\rangle _{{\rm c}}$ is an even function of the magnetic flux enclosed by the compactified dimension. In Fig \ref{fig3} we plot \eqref{azimuthal-current-second-part} as function of $\tilde{\beta}$ for $D=4$, considering $ma=1$, $\xi=0$, $\varepsilon=0.25$ and different values of $q$. As we can see, besides  $\langle j^{\phi}(x)\rangle_{c}$ to present a strong dependence on the parameter $q$, its direction depends on the value of $\tilde{\beta}$. 

In the regime $L/w\gg1$, \eqref{azimuthal-current-second-part} presents the following asymptotic behavior
\begin{eqnarray}
\langle j^{\phi}(x)\rangle_{c}&\approx& \frac{2^{1-2\nu}e\Gamma(D/2+\nu+1)}{(4\pi)^{\frac{D}{2}}\Gamma(\nu+1)a^{D+1}}\bigg(\frac{w}{L}\bigg)^{D+2\nu+2}\sum_{l=1}^{\infty}\cos(2\pi l\tilde{\beta})\nonumber\\
&\times&\Bigg\{\sideset{}{'}\sum_{j=1}^{[q/2]}\sin(2\pi j/q)\sin(2\pi j\varepsilon)
\bigg[\frac{l^2}{4}+\bigg(\frac{r}{L}\bigg)^{2}\sin^2(\pi j/q)\bigg]^{-\frac{D}{2}-\nu-1}
\nonumber\\
&+&\frac{q}{\pi}\int_{0}^{\infty}dy\frac{\sinh{(2y)}g(q,\varepsilon,2y)} {\cosh(2qy)-\cos(\pi q)}\bigg[\frac{l^2}{4}+\bigg(\frac{r}{L}\bigg)^{2} \cosh^2(y)\bigg]^{-\frac{D}{2}-\nu-1}\Bigg\}  \  .
\label{azimuthal-current-second-part-asymptotic}
\end{eqnarray}
For a conformally coupled massless scalar field and taking  $D=4$, a much simpler expression can be provided. It reads,
\begin{eqnarray}
\label{current1}
	\langle j^{\phi}(x)\rangle_{c}&=& \bigg(\frac{w}{aL}\bigg)^{5}\frac{3e}{\pi^2}\Bigg\{\sideset{}{'}\sum_{j=1}^{[q/2]}\sin(2\pi j/q)\sin(2\pi j\varepsilon)\bigg[G_{c}(\tilde{\beta},\rho_{j})-G_{c}(\tilde{\beta},\sigma_{j})\bigg]
	\nonumber\\
	&+&\frac{q}{\pi}\int_{0}^{\infty}dy\frac{\sinh{(2y)}g(q,\varepsilon,2y)}{\cosh(2qy)-\cos(\pi q)}\bigg[G_{c}(\tilde{\beta},\eta(y))-G_{c}(\tilde{\beta},\tau(y))\bigg]\Bigg\} \  ,
\end{eqnarray}
where we have defined the function
\begin{equation}
	G_{c}(\tilde{\beta},x)=\sum_{l=1}^{\infty}\frac{\cos{(2\pi l\tilde{\beta})}}{(l^2+x^2)^{5/2}},
\end{equation}
and introduced new variables
\begin{eqnarray}
	\rho_{j}&=&\frac{2r\sin{(\pi j/q)}}{L}\ \ ,  \quad\quad\quad\quad\quad\quad\quad\eta(y)=\frac{2r\cosh(y)}{L}
	\nonumber\\
	\sigma_{j}&=&\frac{2}{L}\sqrt{w^2+r^2\sin^2{(\pi j/q)}}  \  \ , \quad\quad\quad\tau(y)=\frac{2}{L}\sqrt{w^2+r^2\cosh^2(y)}.
	\label{variables}
\end{eqnarray}
Similarly to what happened with the \eqref{j_conf}, two different contributions appear in \eqref{current1}. The positive contribution is due to the compactification only, and the negative one is induced by the boundary located at $w=0$. Also we can observe that for $r\gg w$ the latter tends to cancel the former. 

\subsection{Current along the compactified dimension}
In this section we want to analyze the current density along the compactified axis, named axial current.  As we shall see, due to the compactification an axial current will be induced. This current goes to zero in the limit $L\to \infty$.
The VEV of the axial current is
calculated by
\begin{equation}
	\langle j_{z}(x)\rangle =ie\lim\limits_{x'\rightarrow x}\{(\partial_{z}-\partial'_{z	})W(x,x')+2ieA_{z}W(x,x')\}
	\label{eqn:axial-current}
\end{equation}
Substituting Eq.\eqref{wight2} into the above expression and using the fact that $A_{z} = -\Phi_{z}/L$, we obtain
\begin{eqnarray}
	\langle j_{z}(x)\rangle&=&-\frac{qea^{1-D}w^{D}} {(2\pi)^{D-3}L}\sum_{n=-\infty}^{\infty}\int d\vec{k} \int_{0}^{\infty}\lambda J^{2}_{q|n+\alpha|}(\lambda r)d\lambda
	\int_{0}^{\infty}p J^{2}_{\nu}(pw)dp\nonumber\\
	&\times&\sum_{l=-\infty}^{\infty}\frac{\tilde{k}_{l}}{\sqrt{\lambda^{2}+p^{2}+\vec{k}^{2}+\tilde{k}_{l}^{2}}} \  ,
\end{eqnarray}
where $\tilde{k}_{l}$ is given by \eqref{const3}.

The sum over the quantum number $l$ is again evaluated by using the Abel-Plana formula given in \eqref{Abel-Plana}. In this case we identify $g(u)=2\pi u/L$, and $f(u)$ is given by \eqref{function}. The first integral on the right hand side is zero due the fact that $g(u)$ is an odd function. Therefore, it remains only the second integral. It reads,
\begin{eqnarray}
	\langle j_{z}(x)\rangle&=&-\frac{2iqea^{1-D}w^{D}} {(2\pi)^{D-2}} \int d\vec{k}
	\int_{0}^{\infty}p J^{2}_{\nu}(pw)dp\sum_{n=-\infty}^{\infty}\int_{0}^{\infty}\lambda J^{2}_{q|n+\alpha|}(\lambda r)d\lambda\nonumber\\
	&\times&\int_{\sqrt{\lambda^2+p^2+\vec{k}^2}}^{\infty}dk_{z}\frac{ k_{z}}{\sqrt{k_{z}^2-\lambda^2-p^2-\vec{k}^2}}\sum_{j=\pm1}^{}\frac{j}{e^{Lk_{z}+2\pi ij\tilde{\beta}}-1}  \  ,
\end{eqnarray}
where we have defined the variable $k_{z}=2\pi u/L$.
Again by using the series expansion, $(e^y-1)=\sum_{l=1}^{\infty}e^{-ly}$, in the above expression, we have
\begin{eqnarray}
		\langle j_{z}(x)\rangle&=&- \frac{4qea^{1-D}w^{D}}{(2\pi)^{D-2}}\sum_{l=1}^{\infty}\sin(2\pi l\tilde{\beta})\int d\vec{k} \int_{0}^{\infty}p J^{2}_{\nu}(pw)dp\sum_{n=-\infty}^{\infty}\int_{0}^{\infty}\lambda J^{2}_{q|n+\alpha|}(\lambda r)d\lambda\nonumber\\
		&\times&\int_{\sqrt{\lambda^2+p^2+\vec{k}^2}}^{\infty}dk_z\frac{ k_ze^{-lLk_z}}{\sqrt{k_z^2-\lambda^2-p^2-\vec{k}^2}}  \  .
		\label{axial-current}
\end{eqnarray}
We can evaluated the integral over $k_z$ with the help of \cite{Grad}, the result is given terms of the Macdonald function of the first order, $K_{1}(z)$. Using the integral representation \eqref{representation-Macdonald} again, and the fact that $K_{\nu}(y)=K_{-\nu}(y)$, we obtain 
\begin{eqnarray}
	\int_{\sqrt{\lambda^2+p^2+\vec{k}^2}}^{\infty}dk_{z}\frac{ k_{z}e^{-lLk_{z}}}{\sqrt{k_{z}^2-\lambda^2-p^2-\vec{k}^2}}=\frac{1}{lL}\int_{0}^{\infty}dte^{-t-(lL)^2(\lambda^2+p^2+\vec{k}^2)/4t}
	\label{integral}
\end{eqnarray}
Substituting \eqref{integral} into \eqref{axial-current}, it is possible to evaluate the integrals over $\lambda$, $p$ and $\vec{k}$, obtaining
\begin{eqnarray}
	\langle j_{z}(x)\rangle&=&-\frac{2qea^{1-D}w^{D}L}{(2\pi)^{D/2}r^{D+2}}\sum_{l=1}^{\infty}l\sin(2\pi l\tilde{\beta})\int_{0}^{\infty}d\chi\chi^{D/2} e^{-\chi[1+(l^2L^2+2w^2)/2r^2]}\nonumber\\
	&\times&I_{\nu}\bigg(\frac{w^2\chi}{r^2}\bigg)\sum_{n=-\infty}^{\infty}I_{q|n+\varepsilon|}(\chi)   \   ,
	\label{axial-current-2}
\end{eqnarray}
where we have introduced the variable $\chi=2tr^2/(lL)^2$. The summation over $n$, can be found in \cite{Braganca:2014qma}, and also given by \eqref{summation-formula} by taking $\Delta\varphi=0$. This sum is:
\begin{eqnarray}
	\sum_{n=-\infty}^{\infty}I_{q|n+\varepsilon|}(\chi)=\frac{e^\chi}{q}-\frac{1}{\pi}\int_{0}^{\infty}dy\frac{e^{-\chi\cosh{(y)}}f(q,\varepsilon,y)}{\cosh{(qy)}-\cos{(\pi q})}+\frac{2}{q}\sideset{}{'}\sum_{k=1}^{[q/2]}\cos{(2\pi k\varepsilon)}e^{\chi\cos{(2\pi k/q)}} \ . 
	\label{summation-formula-2}
\end{eqnarray}
The function, $f(q,\varepsilon,y)$, is defined as
\begin{equation}
	f(q,\varepsilon,y)=\sin{[(1-|\varepsilon|)q\pi]}\cosh(|\varepsilon|qy)+\sin{(|\varepsilon|q\pi)}\cosh{[(1-|\varepsilon|)qy]}.
\end{equation}
Finally substituting \eqref{summation-formula-2} into \eqref{axial-current-2}, we get 
\begin{eqnarray}
\label{axial_1}
	\langle j_{z}(x)\rangle&=&-\frac{2qea^{1-D}w^{D}L} {(2\pi)^{D/2}r^{D+2}}\sum_{l=1}^{\infty}l\sin(2\pi l\tilde{\beta})\int_{0}^{\infty}d\chi\chi^{D/2}e^{-\chi[1+(l^2L^2+2w^2)/2r^2]}I_{\nu}\bigg(\frac{w^2\chi}{r^2}\bigg)\nonumber\\
	&\times&\Bigg[\frac{e^\chi}{q}-\frac{1}{\pi}\int_{0}^{\infty}dy\frac{e^{-\chi\cosh{(y)}}f(q,\varepsilon,y)}{\cosh{(qy)}-\cos{(\pi q})}+\frac{2}{q}\sideset{}{'}\sum_{k=1}^{[q/2]}\cos{(2\pi k\varepsilon)}e^{\chi\cos{(2\pi k/q)}}\Bigg].
\end{eqnarray}

\begin{figure}[!htb]
	\begin{center}
	\includegraphics[scale=0.40]{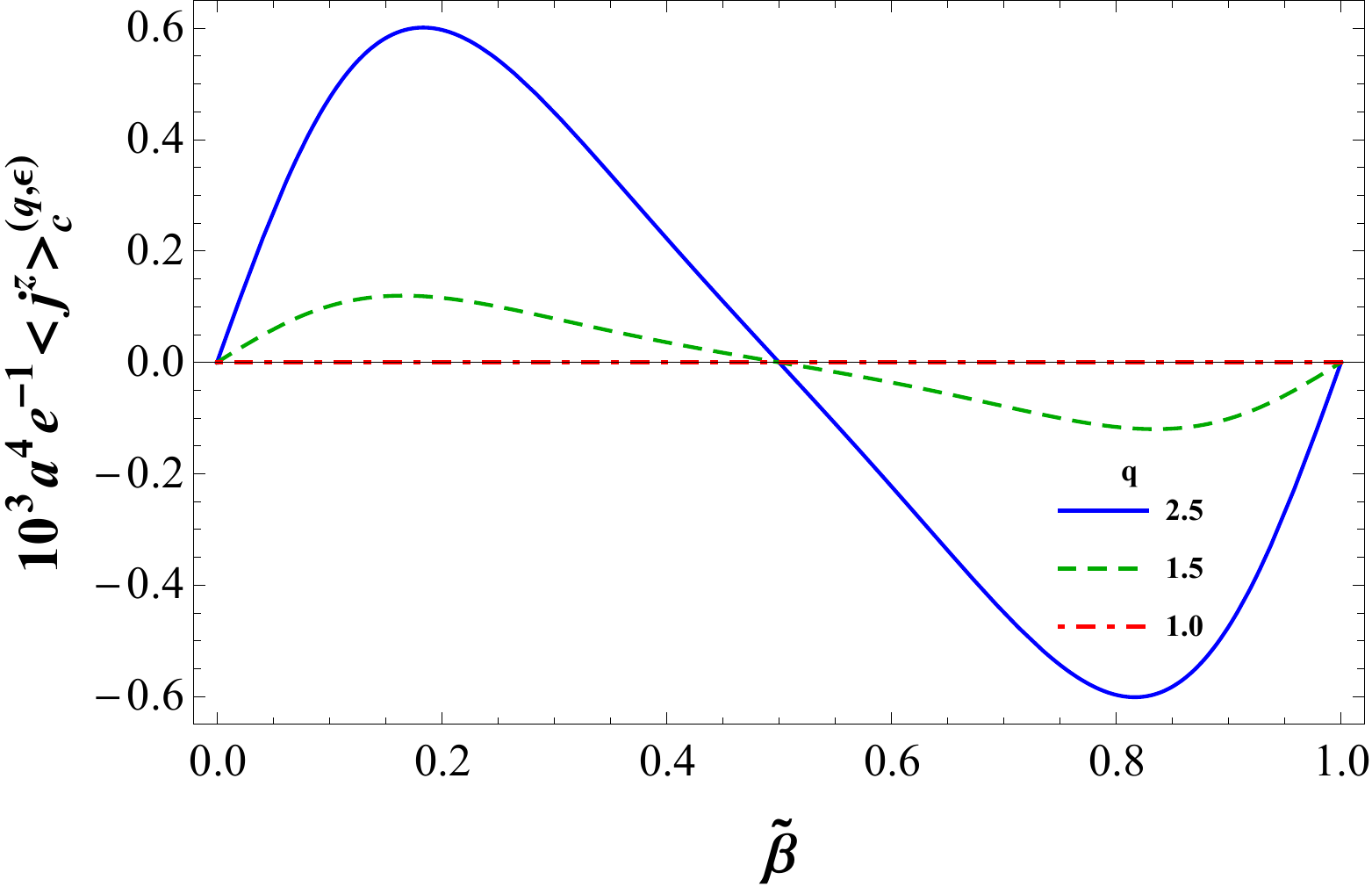}
	\quad
	\includegraphics[scale=0.40]{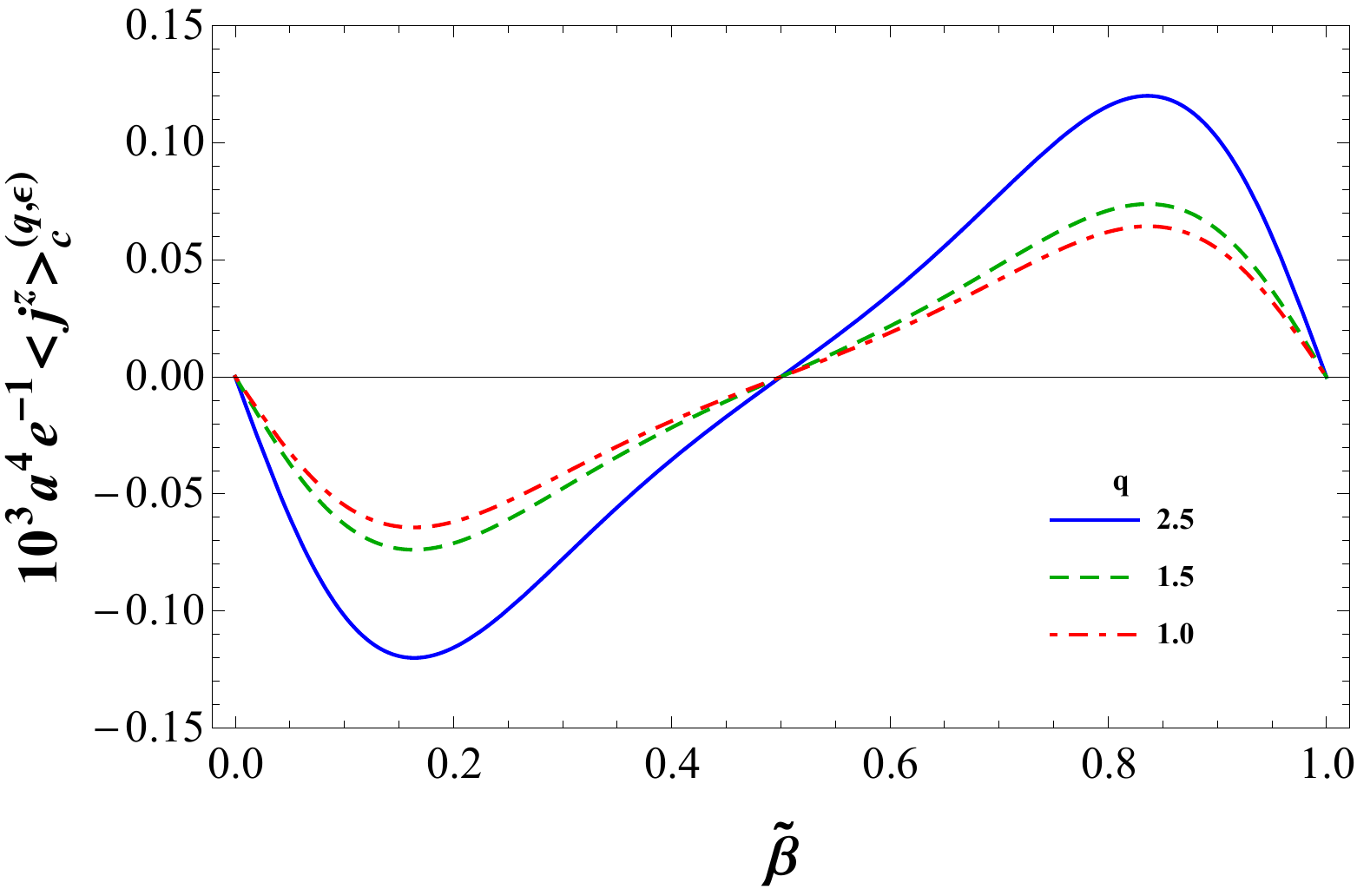}
	\caption{The axial current density is plotted for $D=4$, in units of ``$ea^{-4}$'', as function of $\tilde{\beta}$ for $ma=1$, $\xi=0$, $L/a=1$ and $q=1,1.5$ and $2.5$. In the left plot we consider $\varepsilon=0$, while in the right plot we take $\varepsilon=0.25$.} 
		\label{fig4}
\end{center}
\end{figure}

At this point, we may decompose the current above as
\begin{equation}
	\langle j_{z}(x)\rangle=	\langle j_{z}(x)\rangle_{c}^{(0)}+\langle j_{z}(x)\rangle_{c}^{(q,\varepsilon)}.
	\label{axial-current-decomposition}
\end{equation}
where the first term on the right hand side of the above expression,
\begin{eqnarray}
	\langle j^{z}(x)\rangle_{c}^{(0)}= \frac{4ea^{-(1+D)}L}{(2\pi)^{\frac{D+1}{2}}}\sum_{l=1}^{\infty}l\sin(2\pi l\tilde{\beta})F_{\nu-1/2}^{(D+1)/2}(u_{l0})  \   ,
	\label{topol}
\end{eqnarray}
with $u_{l0}$ given below in \eqref{variables-2}, is purely due to the compactification. It  does not depend on $\varepsilon$ and $q$. For a conformally coupled massless scalar field and taking $D=4$, this contribution reads,
\begin{eqnarray}
	\langle j^{z}(x)\rangle_{c}^{(0)}=\bigg(\frac{w}{a}\bigg)^5\frac{3e} {2\pi^2L^4}\Bigg\{\sum_{l=1}^{\infty}\frac{\sin{(2\pi \tilde{\beta}l)}}{l^{4}}-\sum_{l=1}^{\infty}\frac{l\sin{(2\pi\tilde{\beta}l)}}{\big[l^2+\big(\tiny\frac{2w}{L}\big)^{ \text{\tiny 2}}\big]^{\frac{5}{2}}}\Bigg\} \  .
\end{eqnarray}
The second contribution to the axial current depends on the magnetic fluxes and the parameter associated with the cosmic string. It is given by,
\begin{eqnarray}
\label{axial}
	\langle j^{z}(x)\rangle_{c}^{(q,\varepsilon)}&=& \frac{8ea^{-(1+D)}L}{(2\pi)^{\frac{D+1}{2}}}\sum_{l=1}^{\infty}l\sin{(2\pi l\tilde{\beta})}\Bigg[\sideset{}{'}\sum_{k=1}^{[q/2]}\cos{(2\pi k\varepsilon)}F_{\nu-1/2}^{(D+1)/2}(u_{lk})\nonumber\\
	&-&\frac{q}{\pi}\int_{0}^{\infty}dy\frac{f(q,\varepsilon,2y)}{\cosh{(2qy)}-\cos{(\pi q})}F_{\nu-1/2}^{(D+1)/2}(u_{ly})\Bigg] \  ,
\end{eqnarray}
where we have adopted the following notation,
\begin{eqnarray}
u_{lk}&=&1+\frac{(lL)^2+4r^2\sin^2{(\pi k/q)}}{2w^2}\nonumber\\
u_{ly}&=&1+\frac{(lL)^2+4r^2\cosh^2{(y)}}{2w^2}  \  .
\label{variables-2}
\end{eqnarray}
Notice that this term depends on the radial distance, $r$, and is finite on the string's core. We also can notice that axial current vanishes for integer and half-integer values of $\tilde{\beta}$. In Fig \ref{fig4} we plot the axial current density, Eq.\eqref{axial}, for $D=4$ as function of $\tilde{\beta}$ for $\varepsilon=0$ and $\varepsilon=0.25$, considering $ma=1$, $\xi=0$, $L/a=1$ and different values of $q$. 

We can also analyze the axial current in the regime $L/w\gg1$. Using the asymptotic behavior for the hypergeometric function for large arguments, we get the following expression,
\begin{eqnarray}
\langle j^{z}(x)\rangle_{c}^{(q,\varepsilon)}&\approx& \frac{2^{1-2\nu}e\Gamma(D/2+\nu+1)L}{(4\pi)^{\frac{D}{2}}\Gamma(\nu+1)a^{D+1}}\bigg(\frac{w}{L}\bigg)^{D+2\nu+2}\sum_{l=1}^{\infty}l\sin(2\pi l\tilde{\beta})\nonumber\\
&\times&\Bigg\{\sideset{}{'}\sum_{k=1}^{[q/2]}\cos(2\pi k\varepsilon)
\bigg[\frac{l^2}{4}+\bigg(\frac{r}{L}\bigg)^{2}\sin^2(\pi k/q)\bigg]^{-\frac{D}{2}-\nu-1}
\nonumber\\
&-&\frac{q}{\pi}\int_{0}^{\infty}dy\frac{f(q,\varepsilon,2y)}{\cosh(2qy)-\cos(\pi q)}\bigg[\frac{l^2}{4}+\bigg(\frac{r}{L}\bigg)^{2}\cosh^2(y)\bigg]^{-\frac{D}{2}-\nu-1}\Bigg\}.
\label{current-z-second-part-asymptotic}
\end{eqnarray}
Finally for a conformally coupled massless scalar field and assuming $D=4$, we obtain
\begin{eqnarray}
\label{axial1}
	\langle j^{z}(x)\rangle_{c}^{(q,\varepsilon)}&=& \bigg(\frac{w}{a}\bigg)^5\frac{3e}{\pi^2L^4}\Bigg\{ \sideset{}{'}\sum_{k=1}^{[q/2]}\cos{(2\pi k\varepsilon)}\bigg[V_{c}(\tilde{\beta},\rho_{k})-V_{c}(\tilde{\beta},\sigma_{k})\bigg]\nonumber\\ 
	&-&\frac{q}{\pi}\int_{0}^{\infty}dy\frac{f(q,\varepsilon,2y)}{\cosh{(2qy)}-\cos{(\pi q})}\bigg[V_{c}(\tilde{\beta},\eta(y))-V_{c}(\tilde{\beta},\tau(y))\bigg]\Bigg\}  \  , 
\end{eqnarray}
where we have introduced the function
\begin{equation}
	V_{c}(\tilde{\beta},x)=\sum_{l=1}^{\infty}\frac{l\sin{(2\pi\tilde{\beta}l)}} {(l^2+x^2)^{5/2}} \  , 
\end{equation}
in the integrands of \eqref{axial1}, with the corresponding arguments defined in \eqref{variables}.

\section{Conclude Remarks}
\label{sec4}
In this paper we have investigated the induced scalar current density, $\langle j^\mu\rangle$, in a $(D+1)$- dimensional AdS space, with $D\geq 4$,  admitting the presence of a cosmic string having a magnetic flux running along its axis. In addition we assume the compactification of just one extra dimension in a circle of perimeter $L$ and the existence of a constant vector potential along this direction. This compactification is implemented by assuming that the matter field obeys a quasiperiodicity condition along it, Eq. \eqref{QPC}. In order to develop this analysis we construct the positive energy Wightman function, by solving the Klein-Gordon equation in the corresponding background. By using the Poincar\'e coordinate and admitting a general curvature coupling constant the normalized solution is given by \eqref{COS}. The Wightman function is evaluated by summing over all set of normalized solution \eqref{wight2}. By using the Abel-Plana summation formula, Eq.\eqref{Abel-Plana}, the Wightman function is decomposed in two contributions, one due only to the cosmic string in the AdS background, and the other is induced by the compactification. Fortunately we were able to express this function in a compact form in Eq. \eqref{full-propagator}. 

In our analysis we have proved that only azimuthal and axial current densities are induced.  Due to the compactification, the azimuthal current has been decomposed in two parts. The first one corresponds to the expression in the geometry of a cosmic string in AdS bulk without compactification, and the second is induced by the compactification, both are presented by equations \eqref{azimuthal-current-first-part-3}  and          \eqref{azimuthal-current-second-part}, respectively. Both contributions are odd functions of $\varepsilon$,  with period equal to the quantum flux $\Phi_0$. This is an Aharonov-Bohm-like effect. The pure cosmic string contribution is plotted for $D=4$, in units of the inverse of  $a^{5}e^{-1}$, as function of $\varepsilon$ as shown in Fig.\ref{fig1}. By this graph we can see that the intensity of this current increases with the parameter $q$; also we have plotted  this contribution for two different values of the product $ma$ as function of dimensionless variable $r/w$ for different values of $q$. These graphs are presented in Fig.\ref{fig2}. By them we can see a strong decay in the intensity of $\langle j\rangle_{cs}$; moreover, carefully we can identify that for bigger value of $ma$ the decay is more accentuated. Some asymptotic expressions for this current are provided for specific limiting cases of the physical parameter of the model. For small and larger values of $r/w$, the corresponding asymptotic expressions are given by \eqref{azimuthal-current-first-asymptotic} and   \eqref{azimuthal-current-first-asymptotic-2}, respectively. For $\nu\gg1$ it is given by \eqref{azimuthal-current-first-part-4}. Finally for a conformally coupled massless field, the induced current assume the form  \eqref{j_conf}.

As to the azimuthal current density induced by the compactification, Eq. \eqref{azimuthal-current-second-part}, we can observed that it is an even function of the parameter $\tilde{\beta}$ and is an odd function of the magnetic flux along the core of the string, with period equal to $\Phi_0$. Its dependence on $\tilde{\beta}$ is plotted in Fig.\ref{fig3}, considering the interval $[-0.5, \ 0.5]$ and different values of $q$. Here we also observe that this component depends strongly with $q$. Its asymptotic behavior for large values of $L/w$ is presented in \eqref{azimuthal-current-second-part-asymptotic}, where we can observe that this current  decays with a specific power of $w/L$.

 Due to the compactification, there appears an induced current along the compactified extra dimension presented in a complete expression by \eqref{axial_1}. It has a purely compactification origin and vanishes when $\tilde{\beta}=0,1/2$ and $1$.
 This current can be  expressed as the sum of two terms.  One of them is given by Eq.\eqref{topol}. It is explicit shown that it is independent  of the radial distance $r$, the cosmic string parameter $q$ and $\varepsilon$. The other contribution is given by
 Eq. \eqref{axial}. It depends on the magnetic fluxes  and the planar angle deficit, it is an  odd function of the parameter $\tilde{\beta}$ and is an even 
 function of $\varepsilon$, with period equal to the quantum 
 flux $\Phi_0$. For the particular case when $\beta=0$, 
 Eq. \eqref{axial} becomes an odd function of the magnetic flux enclosed by the compactified dimension.  A plot of the azimuthal current as function ${\tilde{\beta}}$ is presented in Fig.\ref{fig4}  for two different values of $\varepsilon$ and considering $D=4$. By this graph we can see  that the amplitude of the current increases with the parameter $q$ and the effect of $\varepsilon$ is  to change the orientation of the current.

Before to finish this paper, we would like to mention that the currents densities analyzed in this paper refer to the vacuum ones induced by the presence of magnetic fluxes
and the compactification. As was exhibited by all the graphs provided the planar angle deficit associated with the cosmic string spacetime increases the intensity of the azimuthal current density, and the compactification  introduces additional contribution to it; moreover the latter induces a new current density along the compactified dimension.

\section*{Acknowledgments}
 W.O.S thanks CAPES for financial support. H.F.M thanks CNPq. for partial financial support under Grants n$\textsuperscript{\underline{o}}$ 305379/2017-8. E.R.B.M is partially supported by CNPq under Grant n$\textsuperscript{\underline{o}}$ 313137/2014-5.
	
%

\end{document}